\documentclass{article}
\usepackage[utf8]{inputenc}
\usepackage{amsmath}
\usepackage{amssymb}
\usepackage{graphicx}
\usepackage{subcaption}
\usepackage{float}
\usepackage[colorlinks=true, linkcolor=blue, citecolor=blue, urlcolor=blue]{hyperref}
\usepackage{xcolor}
\usepackage[american]{circuitikz}
\usepackage{booktabs}
\usepackage[left]{lineno}
\usepackage{pdfpages}  
\usepackage{scicite}

\usepackage[letterpaper,margin=1in]{geometry}

\date{}

\renewcommand{\thefigure}{S\arabic{figure}}

\renewcommand{\figurename}{Fig.}
\makeatletter
\renewcommand{\fnum@figure}{\textbf{\figurename~\thefigure}}
\makeatother

\makeatletter
\renewcommand{\hyper@natlinkstart}[1]{\hyper@linkstart{link}{page.12}}
\renewcommand{\hyper@natlinkend}{\hyper@linkend}
\makeatother

\makeatletter
\let\orig@thebibliography\thebibliography
\let\orig@endthebibliography\endthebibliography
\renewenvironment{thebibliography}[1]{%
  \setbox0=\vbox\bgroup
  \orig@thebibliography{#1}%
  \setcounter{enumiv}{31}%
}{%
  \orig@endthebibliography
  \egroup
}
\makeatother

\title{ \textbf{Supplementary Materials for:} \\
Cryogenic piezoelectric effects in thin film strontium titanate devices}
\author{Ahmed Khalil, et al.}

\begin{document}
\pdfbookmark{Manuscript}{hiddensection}



# Cryogenic piezoelectric effects in thin film strontium titanate devices

Ahmed Khalil[1,2], Anja Ulrich[1,3], Kamal Brahim[1,4], Andries Boelen[1,5], Danut-Valentin Dinu[1,3], Halil Cuma[1,5], Ioannis Petrides[1], Sandeep Seema Saseendran[1], Xavier Rottenberg[1], Pol Van Dorpe[1,2], Kristiaan De Greve[1,4], Oskar Painter[6,7], Clement Merckling[1,5], Frédéric Peyskens[1], Christian Haffner[1]*

[1] Imec; B-3001 Leuven, Belgium.
[2] Department of Physics and Astronomy, KU Leuven; B-3001 Leuven, Belgium.
[3] Department of Information Technology (INTEC), Photonics Research Group, Ghent University; B-9052 Ghent, Belgium.
[4] Proximus Chair in Quantum Science and Technology, Department of Electrical Engineering (ESAT), KU Leuven; B-3001 Leuven, Belgium.
[5] Department of Materials Engineering (MTM), KU Leuven; B-3001 Leuven, Belgium.
[6] Kavli Nanoscience Institute and Thomas J. Watson, Sr., Laboratory of Applied Physics, California Institute of Technology, Pasadena, CA, USA.
[7] Institute for Quantum Information and Matter, California Institute of Technology, Pasadena, CA, USA.

*Corresponding author. Email: christian.haffner@imec.be

**Abstract**
Next generation quantum technologies will need to rely on the efficient transduction between electrical, optical and mechanical quantum degrees of freedom to generate large-scale entanglement over large distances. The performance of such transducers is fundamentally limited by the cryogenic properties of the underlying materials. Here, we demonstrate that engineering strain in ferroelectric thin-film strontium titanate ($SrTiO_3$) not only results in an exceptionally large Pockels coefficient, but also in a robust linear piezoelectric response at cryogenic temperatures, surpassing previous thin-film benchmarks. We measure piezoelectric tensor elements of $d_{15}$ = 151.8 ± 1.5 pm/V and $d_{33}$ = 54.8 ± 4 pm/V and an effective photoelastic coefficient of $p_{eff}$ =0.56 at 5 Kelvin. Utilizing these enhanced properties, we demonstrate the first $SrTiO_3$-on-oxide acousto-optic modulator with a voltage-length-product ($V_\pi L$) of 0.874 ± 0.084 V cm, outperforming state-of-the-art unreleased modulators that typically feature a $V_\pi L$ of a few V.cm. Our results establish thin-film $SrTiO_3$ as a promising material system for integrated quantum photonics operating at cryogenic temperatures.

**Main**
Hybrid devices are rapidly emerging as pivotal components in the advancement of quantum technologies[1,2]. These hybrid devices facilitate a range of essential functions, from quantum readout of superconducting qubits[3] to acoustic beam steering for ion and neutral atom qubit systems[4], with additional applications as low-loss photonic switches and modulators in optical quantum computing. Among such hybrid devices, popular examples are optical transducers based on opto-mechanical devices, which can be based on capacitive coupling[5] or piezoelectric effects[6]. The latter typically provides the requisite gigahertz bandwidth essential for applications like high-speed qubit readout and precise control of neutral atoms[7]. Piezo-opto-mechanical quantum transducers[8,9], for instance, translate the microwave (μ-wave) information to optical information by utilizing a two-step process based on gigahertz phonons as bosonic intermediary. The piezoelectric effect is used in a first step to convert μ-wave photons to gigahertz phonons, which are subsequently coupled in a second step to an acousto-optic cavity

where the photoelastic effect transduces the phonon to an optical photon. By spatially separating the two processes, small and optimized mode volumes can be utilized in each step to increase the net strength of the interaction, enabling large vacuum coupling rates ($g_0$) up to a megahertz[10–12]. In contrast, direct μ-wave to optical transducer concepts utilize the Pockels effect but must contend with the optical losses when jointly confining μ-wave and optical photons to similar volumes due to light absorption in the metallic μ-wave waveguide. This limits $g_0$ to the order of kilohertz[2,13].

And while parametrically-amplified electro-opto-mechanical systems show promise for achieving ultra-low optical and mechanical losses[5,14], further increases in the electro-mechanical coupling are needed to improve the transducer bandwidth. A global challenge for all microwave-to-optical quantum transducer platforms is to increase the operational bandwidth while limiting the added noise, leading to a bandwidth-noise figure of merit[5]. In the case of piezo-opto-mechanical quantum transducers, one promising avenue consists of improving the material characteristics, including the piezoelectricity, permittivity and (photo)elasticity, as well as sufficiently breaking certain symmetries that otherwise inhibit or reduce these effects[15]. For instance, the material's point group determines whether the piezo-electric response ($d$) is dominated by the linear piezoelectricity (arising from symmetry breaking) or a quadratic electro-strictive mechanism (typical of materials with inversion symmetry.

Linear piezoelectric materials typically chosen for quantum devices, such as aluminium nitride (AlN $|d_{max}|$ ~ 3.9 pm/V) and lithium niobate (LiNbO₃ $|d_{max}|$ ~ 69 pm/V), combine very low losses with moderate piezoelectric coefficients at both ambient and cryogenic temperatures[16]. Recent demonstrations on strontium titanate (SrTiO₃) operated at 4 K reported a bias induced piezoelectric (i.e. electro-strictive effect) and Pockels (i.e. Kerr) coefficient of $d_{max}$ ~ 400 ± 25 pm/V and $r_{eff}$ ~ 1160 ± 30 pm/V, respectively in bulk crystals. This performance enhancement at cryogenic temperatures is enabled by the soft phonon modes in the material that provide a high polarizability, driving the electro-strictive and Pockels coefficient[17,18]. Integrating SrTiO₃ into large-scale photonics platforms based on epitaxially grown thin films resulted in an effective Pockels coefficient of 345 ± 10 pm/V beyond that of conventional thin films[19].

Here, we show that ferroelectric thin-film SrTiO₃ on SiO₂ not only exhibits a large Pockels coefficient but also features a large linear piezoelectric effect, which contrary to the bias induced piezoelectric effect in bulk SrTiO₃ is maintained at zero bias. The measured material piezoelectric tensor elements $d_{15}$ = 151.8 ± 1.5 pm/V, and $d_{33}$ of 54.8 ± 4 pm/V, exceed previous values for thin-film materials at cryogenic temperatures. Using these effects, we demonstrate here the first SrTiO₃ on oxide acousto-optic modulator - featuring a $V_\pi L$ product of 0.874 ± 0.084 V cm which goes well beyond state-of-the-art: typical unreleased (i.e. not-suspended) acousto-optic modulators feature a $V_\pi L$ on the order of a few V cm or more[16]. As the photoelastic interaction takes place in the thin-film SrTiO₃, we benefit from the material's stronger photoelastic coefficient of $p_{eff}$=0.56 compared to lithium niobate's $p_{max}$ = 0.18[20].

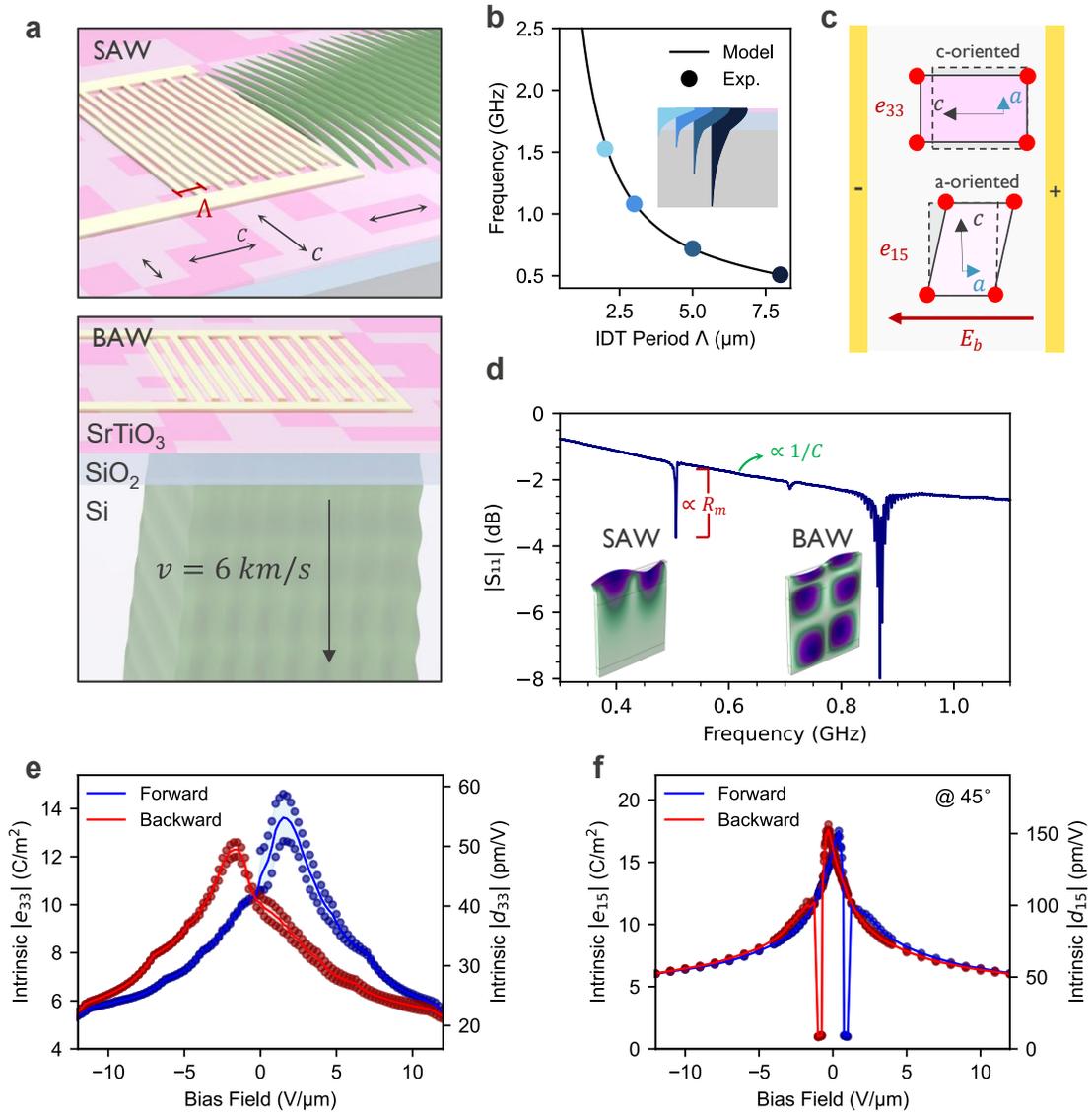

**Figure 1** Experimental quantification of piezoelectric tensor elements in thin-film SrTiO₃ at 5 Kelvin. (a) Artistic schematic of interdigitated transducers (IDTs) launching surface (SAW) and bulk (BAW) acoustic waves. Domain sizes are enlarged for visualization purposes. (b) SAW excitation frequency for varying IDT periodicity (line = simulation, dot = experiment). (c) Domain configurations and polarization directions with respect to the bias electrical field of the IDT. (d) Electrical reflection ($S_{11}$) spectra reveal distinct SAW and BAW resonances, attributed to $e_{33}$ and $e_{15}$ tensor elements of an 8 μm pitched IDT. The BAW's megahertz free-spectral range is a result of the standing acoustic wave formation in the silicon substrate. (e-f) Voltage-dependent piezoelectric coefficients display ferroelectric hysteresis, with $e_{15}$ showing a clear zero crossing at the coercive field ($E_c \sim 0.75$ V.μm⁻¹), while $e_{33}$ remains offset.

To quantify the cryogenic piezoelectric tensor of a-axis (i.e. polarization axis in plane) thin-film SrTiO₃[21], we fabricated interdigitated transducers (IDTs) on oxide-bonded strontium titanate. The heterostructure stack consists of a 200 mm process that bonds 105 nm of SrTiO₃ onto a 2 μm SiO₂/725 μm Si wafer[19]. The measured permittivity response for both 0º and 90º oriented IDTs exhibited identical behavior – suggesting, instead of single domain behavior, rather some mixture of domains. In our analysis we assume a 90º twin boundary domain film with equal populations. This interpretation would align with observations in previous studies on the Pockels coefficient of SrTiO₃ and BaTiO₃ grown on silicon by MBE [22,23]. The IDTs consist of a periodic array of metallic fingers deploying an alternating voltage on a piezoelectric

3

film, thereby launching surface acoustic (SAW) and bulk acoustic (BAW) waves (SAW: Rayleigh mode and BAW: bulk shear mode), see Figure 1(a). The frequencies of these two modes are determined by the sound velocity in the multilayer stack, the periodicity of the fingers and modal dispersion. Literature suggests a slow acoustic velocity in SrTiO$_3$ ($v_{STO} \approx$ 1100 $m/s$)[24]. This is confirmed by comparing measured excitation frequency for different IDT periodicities with the numerically calculated excitation frequency using reported literature values of low-temperature elasticity[25,26], see Figure 1(b). The nonlinear increase for shorter periodicities results from the stronger confinement of the acoustic field to the low-velocity SrTiO$_3$, resulting in a larger excitation frequency.

We extracted the voltage-dependent piezoelectric tensor elements by analyzing the electrically biased spectral response of the electrical reflection coefficient ($S_{11}$). Figure 1(d) shows $S_{11}$ under a biasing field of 3 V/μm, and two narrowband drops in reflected power can be observed at 504 MHz and 874 MHz, respectively caused by the transduction of electromagnetic waves into acoustic waves – a process that acts as a net absorption mechanism and therefore reduction in reflected power. By solving for the SAW and BAW eigenfrequencies of the stack and assuming a crystallographic point group P4mm (tetragonal) for SrTiO$_3$[27], we can associate the first dip (504 MHz) with a SAW Rayleigh wave driven by $e_{33}$ of c-oriented domains, and the second (874 MHz) with a BAW shear wave corresponding with the $e_{15}$ piezoelectric stress coefficient of the a-oriented domains (Fig. 1.c)- see Supplementary Note 1.4. The extracted single-domain piezoelectric $e_{33}$ and $e_{15}$ coefficients are shown in Figure 1(e) and Figure 1(f) while sweeping a bias field forward (blue) and backward (red), respectively. The hysteretic behavior is consistent with ferroelectricity, whereas the non-zero response at zero bias indicates that the linear piezoelectric effect dominates over the electro-strictive effect at 5 Kelvin. Further indications of ferro-electricity are the fact that $e_{15}$ exhibits a zero crossing at the coercive field (E$_C$) of ~ 0.75 V/μm, matching previous results on Pockels measurements[19]. Somewhat surprisingly, $e_{33}$ does not show this behavior an effect that is at present still under investigation. The strength of these tensor elements reaches values up to $e_{33}$ = 13.63 ± 0.98 C.m$^{-2}$ and $e_{15}$ = 17.59 ± 0.4 C.m$^{-2}$ under low biasing fields. Using literature values for the elasticity (*c*) of the shear wave ($c_{44}$ = 115 GPa) and longitudinal wave ($c_{33}$ = 337 GPa) in strontium titanate[26] results in large piezoelectric strain coefficients $d_{15}$ = 151.8 ± 1.5 pm/V, and $d_{33}$=54.8 ± 4 pm/V.

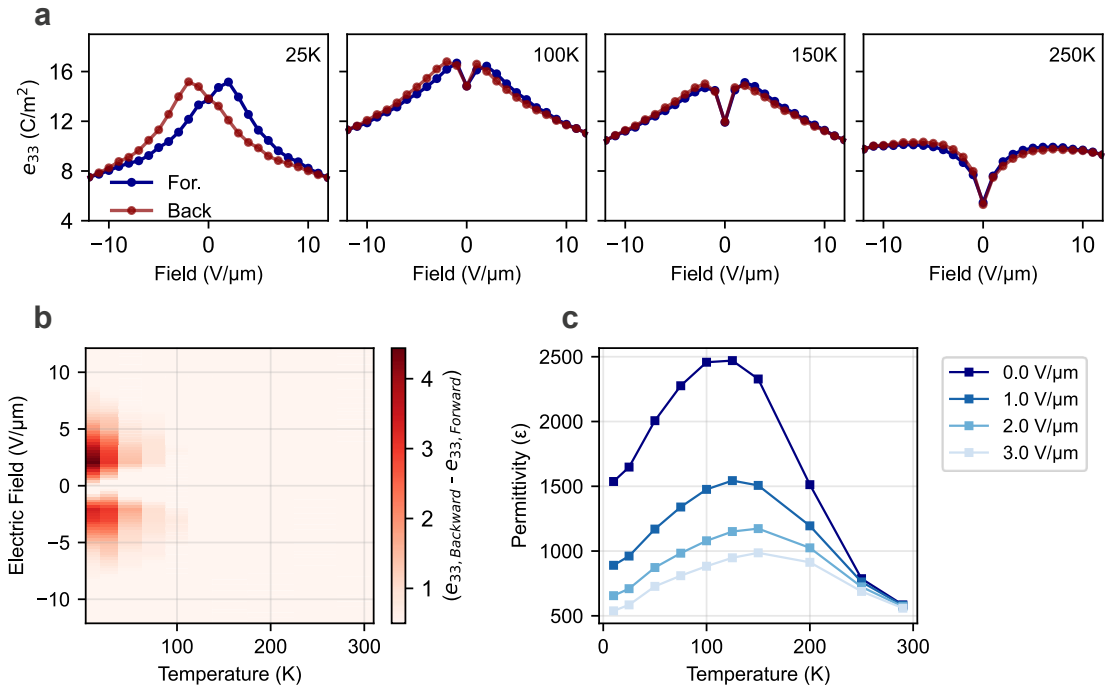

Figure 2 a) *Evolution of the piezoelectric stress coefficient as a function of electric field and temperature highlighting the transition from linear to electro-strictive behaviour at higher temperatures.* b) *At low temperatures, the piezoelectric stress coefficient exhibits pronounced hysteresis, indicative of ferroelectric domain switching.* c) *Permittivity versus temperatures under various bias conditions indicating a voltage-dependent ferroelectric-to-paraelectric phase transition around 100K to 150 K.*

Figure 2a) illustrates the evolution of the piezoelectric response as a function of electric field and temperature, highlighting the transition from a linear piezoelectric regime to electrostrictive behavior. At low temperatures (e.g. 10 K), the piezoelectric coefficient exhibits pronounced hysteresis between forward and backward sweeps, along with a non-zero value at zero bias. This behavior is indicative of a linear piezoelectric drive mechanism, likely associated with ferroelectric domain switching. Above $T_c$ (~100 K) the piezoelectric coefficient shows a drop at zero bias, signaling the presence of electrostriction in the paraelectric phase. With increasing temperature the hysteresis progressively diminishes (Figure 2b). Rather than a sharp transition, the ferroelectric–paraelectric phase change extends over a broad temperature range (100–150 K), as shown in Fig. 2c. This diffusive transition is indicative of relaxor ferroelectric behaviour, typically linked to the presence of polar nanoregions that smear the response around the Curie temperature[28].

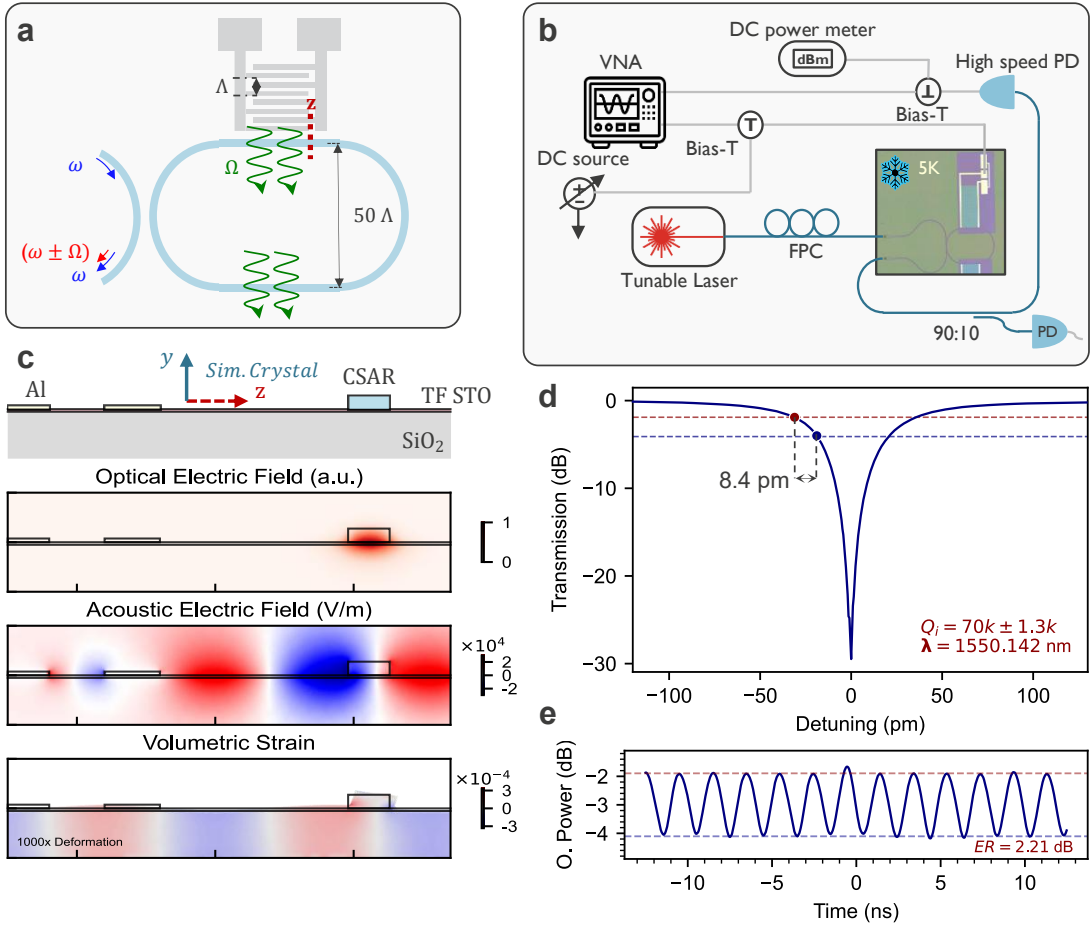

Figure 3 a) Schematic of the acousto-optic modulator. Light in a bus waveguide (ω) couples to a racetrack ring resonator and is modulated by an IDT operated at Ω resulting in a modulated output signal ω±Ω. The multiple integer relation between the IDT's periodicity and resonator diameter is chosen to support a push-push modulation of the two racetracks' arms. The cross section at the red dashed line is shown in c). b) Opto-electro-mechanical setup used to characterize acousto-optic modulation. c) Schematic cross-section of the device stack. The simulated IDT includes the complete set of finger pairs (N = 60) and is shown together with the optical mode, acoustically induced electric field, and volumetric strain profiles in the waveguide cross-section, highlighting contributions from photoelasticity, moving-boundary effects, and a weak secondary electro-optic response. d) Optical power transmission in the bus waveguide coupled to the racetrack resonator shows a Q ~ 70k ± 1.3k corresponding to a modal propagation loss of ~4.7dB/cm. e) Driving the device under 504 MHz and 235 mV peak voltage results in an optical power modulation of 2.2 dB translating to a $V_\pi L$ of 0.874 V.cm.

The observation of the large piezoelectric effect, in combination with large Pockels effects observed in these layer stacks, suggests the presence of potentially large photoelasticity as well. To investigate the latter, we designed devices that leverage photoelasticity to enable acousto-optic modulation. Specifically, we use an IDT to generate acoustic waves that mediate acousto-optic conversion in a critically coupled optical racetrack resonator at 1550 nm (free-spectral range 0.781 nm), see Figure 3a. The resonator comprises of a 550 nm high and 1500 nm wide single-mode CSAR waveguide (E-Beam resist) on top of the thin-film $SrTiO_3$, thereby utilizing large photoelasticity in the complex oxide, see Figure 3c. The IDT launched waves under 90º with respect to the 200 μm long racetrack sections. Due to the negligible decay of the acoustic wave, the light experiences the same phase modulation a second time in the opposite arm resulting in a push-push operation of the resonator (Supplement 1.6). Due to the multi-domain nature of the film, the utilized piezoelectric coefficient of the acoustic wave is two times weaker compared to the single domain coefficient due to averaging over domains that are parallel (c -

6

contributing) and perpendicular (a - not contributing) to the electrical field within the IDT. The measured resonator linewidth of the fundamental TE mode is $\kappa/2\pi=5.63 \pm 0.07$ GHz (Figure 3d), corresponding to modal propagation losses of ~4.68 dB/cm. , The device was operated in the non-resolved sideband regime as the mechanical excitation frequency of 504 MHz is smaller than the linewidth of the optical cavity. To converted the acousto-optic phase modulation into an amplitude modulation we detuned the laser to the 3 dB transmission point of the resonator (i.e. $f_{res} + \kappa/2$) . Figure 3e) shows a 2.2 dB modulation of the optical transmitted power when applying a 504 MHz electrical signal with a peak voltage of 235 mV. This translates to a resonance shift of 8.4 pm, yielding a $V_\pi L$ of $0.874 \pm 0.084$ V cm based on the 780 pm FSR.

The acousto-optic modulation in such devices can originate from three distinct physical mechanisms: the photoelastic effect, the moving-boundary effect and a secondary electro-optic (Pockels) effect. The photoelastic effect mechanically strains the material at an atomic level which modifies its refractive index through the material's photoelastic tensor. The moving boundary effect does not change a material's refractive index itself but, for instance, physically displaces the CSAR such that the optical mode experiences a different electromagnetic environment such as vacuum that features a lower refractive index. Finally, the secondary electro-optical effect originates from the acoustic wave inducing a local electric field via the inverse piezoelectric effect. This electric field interacts with the optical mode through the material's Pockels tensor, leading to an additional modulation component. Figure 3c) depicts the waveguide and IDT cross-section on top, followed by the simulated waveguide confined optical field, acoustic-induced electrical field and volumetric strain. From our numerical analysis we find that the overall phase shift is primarily governed by the photoelastic effect (>90%). The contribution of moving boundaries is limited by the not suspended waveguide. The secondary electro-optic effect is not contributing as the waveguide is oriented parallel to the weakest Pockels coefficient of $SrTiO_3$ $r_{33}$ ~ 40 pm/V[29] and the domains under the waveguide are randomly oriented as the piezoelectric induced field of < 0.1 V um$^{-1}$ is below the coercive field.

Despite the absence of significant secondary contributions, we report a low voltage-length product driven by the large piezoelectric coefficient in combination with the high photoelastic coefficient of $p_{eff} = 0.56$ in $SrTiO_3$. The latter being three times larger compared to lithium niobate's photo elasticity. [Supplementary Note 2.2].

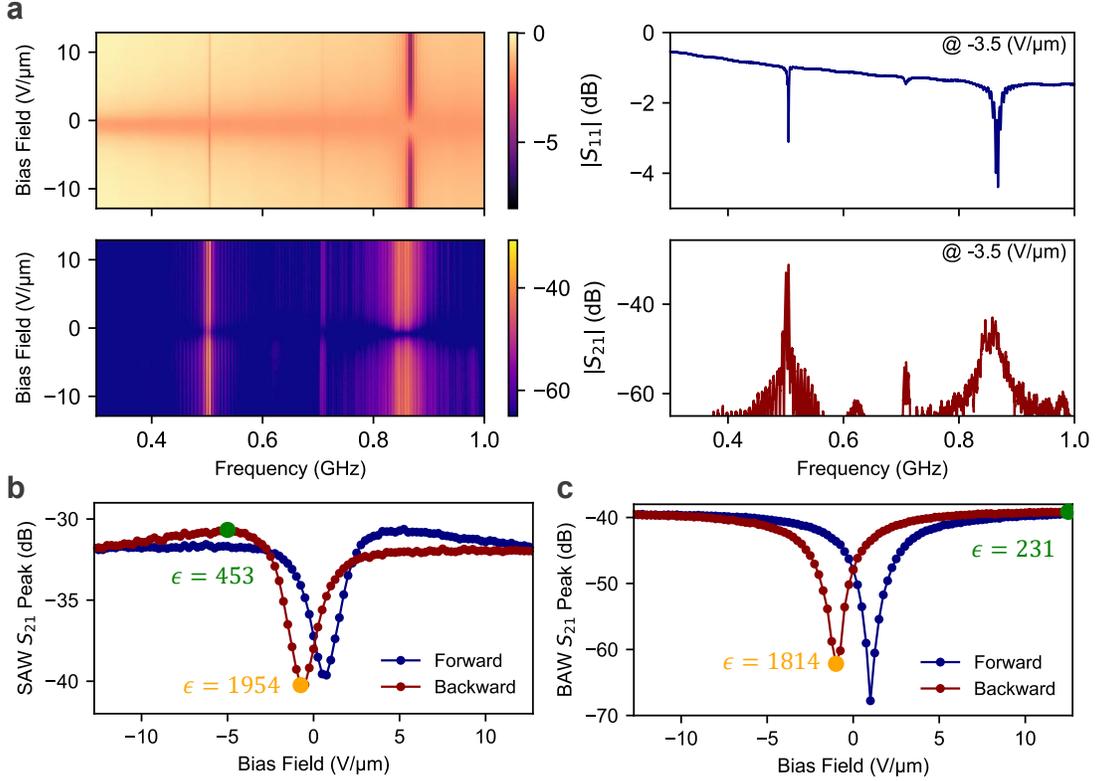

Figure 4 a) Electro-acoustic $S_{11}$ and acousto-optic $S_{21}$ spectra measured during a downward bias sweep, showing the correlation between piezoelectric-driven acoustic excitation and optical phase modulation. The line traces show a cutline under a $-3.5$ V μm$^{-1}$ bias, reaching a maximum $S_{21}$ of $-30.5$ dB. The increase of $S_{21}$ in (b) SAW and (c) BAW modes for fields beyond the coercive field arises from improved impedance matching due to the electric-field-induced reduction of the permittivity of SrTiO$_3$.

Ferroelectric materials operated close to their Curie temperature are known for their high sensitivity to an applied electric field[30]. Thus, upon applying a bias field to the IDT one can change the permittivity of the thin-film from >2000 to ~500, see Figure 2c). This results in a large tunability of the device's capacitances and thus the impedance matching of the device can be controlled to maximize the electrical to acoustic power transduction. Figure 4a) shows the electro-acoustic ($S_{11}$) and acousto-optical ($S_{21}$) interaction versus frequency for a downward bias voltage sweep. A clear correlation between electro-acoustic transduction and acousto-optical modulation is visible. While the piezoelectric coefficient $e_{33}$ reported in Figure 1e) peaks at ~1.5 V.μm$^{-1}$ before dropping by 50%, both $S_{11}$ and $S_{21}$ continue to increase with increasing voltage, see Figure 4b,c, due to the improved impedance matching between the RF source and the transducer. This allows for optimization of the acousto-optic conversion efficiency under varying operational conditions[31].

**Conclusion**

In this work, we have, for the first time, demonstrated pronounced piezoelectric and photoelastic effects in epitaxial thin-film SrTiO$_3$ at cryogenic temperatures, achieving modulation efficiencies that exceed previous devices of similar architecture. We assume that our device's performance is currently limited by the presence of 90º twin boundaries within the film. This multi-domain structure halves the effective utilization of the extracted material coefficients $e_{15}$ and $e_{33}$, resulting in a doubling of the V$_\pi$L compared to what could be realized with a single-domain film. With further improvements in film quality and domain engineering, particularly by reducing or eliminating twin boundaries, the intrinsic material coefficients

8

could approach those found in bulk, single-domain $SrTiO_3$. This work opens new opportunities for scalable, high-performance quantum and classical photonic systems by unlocking the remarkable capabilities of $SrTiO_3$.

**Contributions:**
A.K. designed the IDTs and the acousto-optic modulator, performed the measurements and data analysis. A.U. fabricated the samples, built the experimental setup, and assisted with RF, acousto-optic, and temperature-dependent measurements and analysis. K.B. assisted with RF measurements and analysis. A.B. epitaxially grew $SrTiO_3$ and performed wafer characterization. D.D. analysed the photoelastic response and assisted with acousto-optic simulations. H.C. assisted the fabrication and optical loss analysis. S.S.S. developed the wafer-bonding process in the fab. I.P. contributed to the conceptual development of the work. X.R., P.V.D., O.P., C.M., F.P. and C.H. conceived and supervised the project. A.K., I.P., O.P., K.D.G., D.D., F.P. and C.H. prepared the manuscript, with input from all authors.

**Competing interests** The authors declare no competing interests.
**Data and materials availability** All data are available in the main text or the supplementary materials. Requests for further materials should be addressed to Christian Haffner.

# Methods

**Growth**
Epitaxial SrTiO₃ thin films were grown on p-type Si (001) substrates by molecular beam epitaxy. After a 2% HF cleaning, the substrates were thermally annealed in vacuum and underwent a Sr-assisted native oxide desorption. A ½-monolayer Sr was grown prior to SrTiO₃ growth as an oxidation barrier, as confirmed by a (2 × 1) surface reconstruction.
The initial SrTiO₃ layers (~3 nm) were grown at 350 °C under molecular oxygen, followed by continued growth at 550 °C under atomic oxygen. The films exhibit an epitaxial relationship with a 45° in-plane rotation relative to the Si substrate ([100] SrTiO₃ ∥ [110] Si) and were grown at a rate of approximately 1 nm min$^{-1}$.

**Wafer Bonding**
Following growth, the SrTiO₃/Si wafers were annealed in oxygen at 850 °C for 30 min to reduce oxygen vacancy concentration induced during low-pressure growth. A SiO₂ layer was then deposited by chemical vapor deposition to enable oxide–oxide wafer bonding to a SiO₂-on-Si wafer. After bonding, the original Si substrate was removed by mechanical grinding followed by selective wet etching, yielding a final stack consisting of 105 nm SrTiO₃ on 2 μm SiO₂ on 725 μm Si. A second oxygen anneal at 850 °C for 30 min was performed to restore oxygen stoichiometry and improve crystallinity after substrate removal.

**Device Fabrication**
Electron-beam lithography (EBL) was used to pattern a bilayer resist stack defining the interdigitated transducer (IDT) geometry. Following development, a 200 nm aluminium layer was deposited using electron-beam evaporation and subsequently structured through liftoff. Next, a 550 nm thick CSAR 62 resist was spin-coated and patterned by electron-beam lithography, defining optical waveguides 1.5 μm in width with a minimum bending radius of 200 μm.

**Experimental setup**
The reflection spectrum of the IDTs was measured in an Attocube attoDRY800 cryostat at 5 K. Electrical access to the IDTs was achieved using a ground–signal (GS) probe connected to a (Keysight P5003B) vector network analyzer (VNA), operated at an output power of 0 dBm (corresponding to an electric field of 0.15 V.μm$^{-1}$). A bias tee was inserted in the probe path to apply a DC bias from a source/measure unit (Keysight B2910BL). Optical access to the device was enabled using a fiber array aligned to the on-chip grating couplers. The transmitted optical signal was divided using a 90:10 splitter: the weaker arm was used to monitor the optical operation point, while the main arm was directed to a high-speed, DC-coupled photoreceiver (Newport 1544-B) with a conversion gain of 800 V.W$^{-1}$. The electrical output of the photoreceiver was further separated using a bias tee, with the DC component routed to a power meter and the AC component connected to the VNA for RF measurements. Under these conditions, a peak transmission coefficient of $S_{21}$ = - 30.5 dB was measured, with an optical power of -19 dBm incident on the high-speed photoreceiver. In the $V_\pi L$ measurements, the output of the photodetector was directly connected to the oscilloscope while being driven by 504 MHz signal with 1 dBm RF power.

**Permittivity extraction**
The imaginary part of the extracted IDT impedance, Z(f), over a broad frequency range away from resonance is dominated by the IDT static capacitance in series with a parasitic inductance as

$$Im[Z(f)] = \frac{1}{2\pi f C_T} + 2\pi f L_s. \tag{1}$$

To determine the static capacitance $C_T$, we fit the $Im[Z]$ across a 1 GHz frequency span. This method is justified because the piezoelectric contribution to $Im[Z(f)]$ is negligible over a wide frequency range. To extract the SrTiO$_3$ permittivity $\varepsilon_{STO}$, we use the electrostatics module of COMSOL Multiphysics to map the measured static capacitance ($C_T$) to the dielectric permittivity of the SrTiO$_3$ thin film. Extended Data Fig. 1a, shows the extracted permittivity as a function of the applied bias field.

Due to the multidomain nature of the SrTiO$_3$ film, the effective permittivity does not depend on the in-plane orientation of the IDT fingers. Analogous to the refractive-index ellipse, the angular dependence of the dielectric permittivity is described by an ellipse whose semi-axes are set by the principal permittivities along each crystal axis[22]

$$\begin{aligned}\varepsilon_{\text{eff}}(\theta) &= \frac{1}{2}\left[\underbrace{\varepsilon_c \sin^2\theta + \varepsilon_a \cos^2\theta}_{\text{a-oriented}} + \underbrace{\varepsilon_a \sin^2\theta + \varepsilon_c \cos^2\theta}_{\text{c-oriented}}\right] \\ &= \frac{\varepsilon_a + \varepsilon_c}{2}[\sin^2\theta + \cos^2\theta] \\ &= \frac{\varepsilon_a + \varepsilon_c}{2},\end{aligned} \tag{2}$$

where $\varepsilon_a$ and $\varepsilon_c$ are the permittivities along the a- and c-axes, respectively.

**IDTs with different angles**

Due to the multidomain nature of the SrTiO$_3$ thin films, the measured effective piezoelectric response was identical for both 0° and 90° IDTs, as shown in Extended Data Fig. 1b. The only difference between the two orientations was the parasitic resistance $R_s$ which can be attributed to the different electrode configuration used to ease the measurement process (See SI. Fig. 1). This observation further confirms that the domains are randomly oriented in-plane with equal probability along the a- and c- axes.

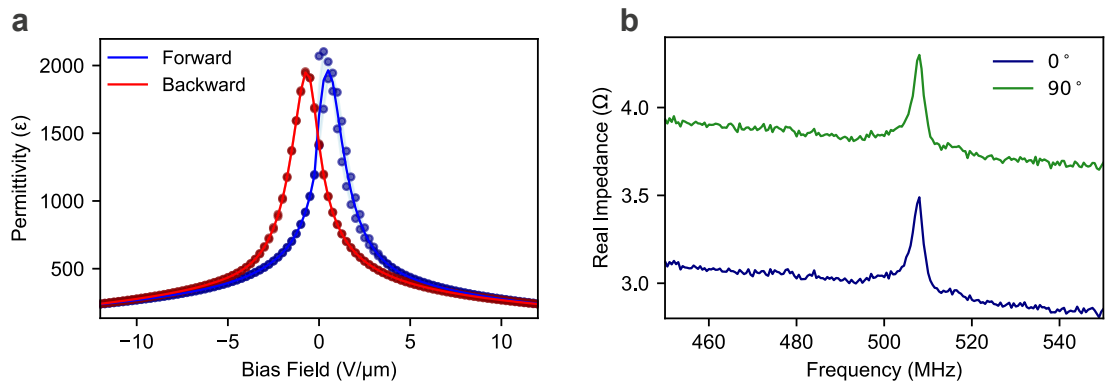

Extended Data Figure 1. a) Effective permittivity of SrTiO$_3$ as a function of the applied DC bias field at 5 K, measured over a frequency range up to 1 GHz. b) Real part of the impedance for IDTs oriented at 0° and 90°, showing identical electromechanical response apart from a parasitic resistance offset arising from electrode routing. The orientation independence reflects the multidomain nature of the SrTiO$_3$ thin film.

**References**

1. Sekine, A., Murakami, R. & Doi, Y. Microwave-to-Optical Quantum Transduction of Photons for Quantum Interconnects. Preprint at https://doi.org/10.48550/arXiv.2509.26349 (2025).

2. Coherent control of a superconducting qubit using light | Nature Physics. https://www.nature.com/articles/s41567-025-02812-0.

3. Mirhosseini, M., Sipahigil, A., Kalaee, M. & Painter, O. Superconducting qubit to optical photon transduction. *Nature* **588**, 599–603 (2020).

4. Li, B., Lin, Q. & Li, M. Frequency–angular resolving LiDAR using chip-scale acousto-optic beam steering. *Nature* **620**, 316–322 (2023).

5. Zhao, H., Chen, W. D., Kejriwal, A. & Mirhosseini, M. Quantum-enabled microwave-to-optical transduction via silicon nanomechanics. *Nat. Nanotechnol.* **20**, 602–608 (2025).

6. Forsch, M. *et al.* Microwave-to-optics conversion using a mechanical oscillator in its quantum ground state. *Nat. Phys.* **16**, 69–74 (2020).

7. Zhao, M. *et al.* An integrated photonics platform for high-speed, ultrahigh-extinction, many-channel quantum control. Preprint at https://doi.org/10.48550/arXiv.2508.09920 (2025).

8. Weaver, M. J. *et al.* An integrated microwave-to-optics interface for scalable quantum computing. *Nat. Nanotechnol.* **19**, 166–172 (2024).

9. Jiang, W. *et al.* Optically heralded microwave photon addition. *Nat. Phys.* **19**, 1423–1428 (2023).

10. MacCabe, G. S. *et al.* Nano-acoustic resonator with ultralong phonon lifetime. *Science* **370**, 840–843 (2020).

11. Balram, K. C., Davanço, M. I., Song, J. D. & Srinivasan, K. Coherent coupling between radiofrequency, optical and acoustic waves in piezo-optomechanical circuits. *Nat. Photonics* **10**, 346–352 (2016).

12. Balram, K. C., Davanço, M. I., Song, J. D. & Srinivasan, K. Coherent coupling between radiofrequency, optical and acoustic waves in piezo-optomechanical circuits. *Nat. Photonics* **10**, 346–352 (2016).

13. Lecocq, F. *et al.* Control and readout of a superconducting qubit using a photonic link. *Nature* **591**, 575–579 (2021).

14. Delaney, R. D. *et al.* Superconducting-qubit readout via low-backaction electro-optic transduction. *Nature* **606**, 489–493 (2022).

15. Dielectric Phenomena in Solids. *ScienceDirect* http://www.sciencedirect.com:5070/book/monograph/9780123965615/dielectric-phenomena-in-solids.

16. Tian, H. *et al.* Piezoelectric actuation for integrated photonics. *Adv. Opt. Photonics* **16**, 749 (2024).

17. Anderson, C. P. *et al.* Quantum critical electro-optic and piezo-electric nonlinearities. *Science* **390**, 394–399 (2025).

18. Grupp, D. E. & Goldman, A. M. Giant Piezoelectric Effect in Strontium Titanate at Cryogenic Temperatures. *Science* **276**, 392–394 (1997).

19. Ulrich, A. *et al.* Engineering high Pockels coefficients in thin-film strontium titanate for cryogenic quantum electro-optic applications. *Science* **390**, 390–393 (2025).

20. Dixon, R. W. Photoelastic Properties of Selected Materials and Their Relevance for Applications to Acoustic Light Modulators and Scanners. *J. Appl. Phys.* **38**, 5149–5153 (1967).

21. Boelen, A. *et al.* Stoichiometry and Thickness of Epitaxial SrTiO3 on Silicon (001): An Investigation of Physical, Optical, and Electrical Properties. *Cryst. Growth Des.* **25**, 5752–5760 (2025).

22. Barium titanate and lithium niobate permittivity and Pockels coefficients from megahertz to sub-terahertz frequencies | Nature Materials. https://www.nature.com/articles/s41563-025-02158-1.

23. Zhang, P. *et al.* Intrinsic-strain-induced ferroelectric order and ultrafine nanodomains in SrTiO3. *Proc. Natl. Acad. Sci. U. S. A.* **121**, e2400568121 (2024).

24. Yang, D. *et al.* Surface acoustic wave generation and detection in the quantum paraelectric regime of SrTiO$_3$-based heterostructures. *Phys. Rev. B* **108**, L041402 (2023).

25. Bell, R. O. Elastic Constants of Strontium Titanate. *Phys. Rev.* **129**, 90–94 (1963).

26. Scott, J. F. & Ledbetter, H. Interpretation of elastic anomalies in SrTiO3 at 37K. *Z. Für Phys. B Condens. Matter* **104**, 635–639 (1997).

27. Haeni, J. H. *et al.* Room-temperature ferroelectricity in strained SrTiO3. *Nature* **430**, 758–761 (2004).

28. Damjanovic, D. Ferroelectric, dielectric and piezoelectric properties of ferroelectric thin films and ceramics. *Rep. Prog. Phys.* **61**, 1267 (1998).

29. Ulrich, A. High speed ferroelectric tensor of thin-film strontium titanate.(Manuscript in preparation).

30. Eggli, R. S. *et al.* Cryogenic hyperabrupt strontium titanate varactors for sensitive reflectometry of quantum dots. *Phys. Rev. Appl.* **20**, 054056 (2023).

31. Electrostriction: Nonlinear Electromechanical Coupling in Solid Dielectrics | The Journal of Physical Chemistry B. https://pubs-acs-org.kuleuven.e-bronnen.be/doi/10.1021/jp971522c.



\begin{center}
\section*{Supplementary Materials for\\ Cryogenic piezoelectric effects in thin film \\strontium titanate devices}

Ahmed Khalil et al.\\ 
\small$^\ast$Corresponding author. Email: Christian.Haffner@imec.be\\
\subsubsection*{This PDF file includes:}
Supplementary Text\\
Figures S1 to S9\\
Tables S1 to S2\\
\end{center}

\newpage
\section{Piezoelectricity extraction}

\begin{figure}[h]
    \centering
    \resizebox{0.45\textwidth}{!}{%
    \begin{subfigure}[b]{0.25\textwidth}
        \centering
        \includegraphics[width=\textwidth]{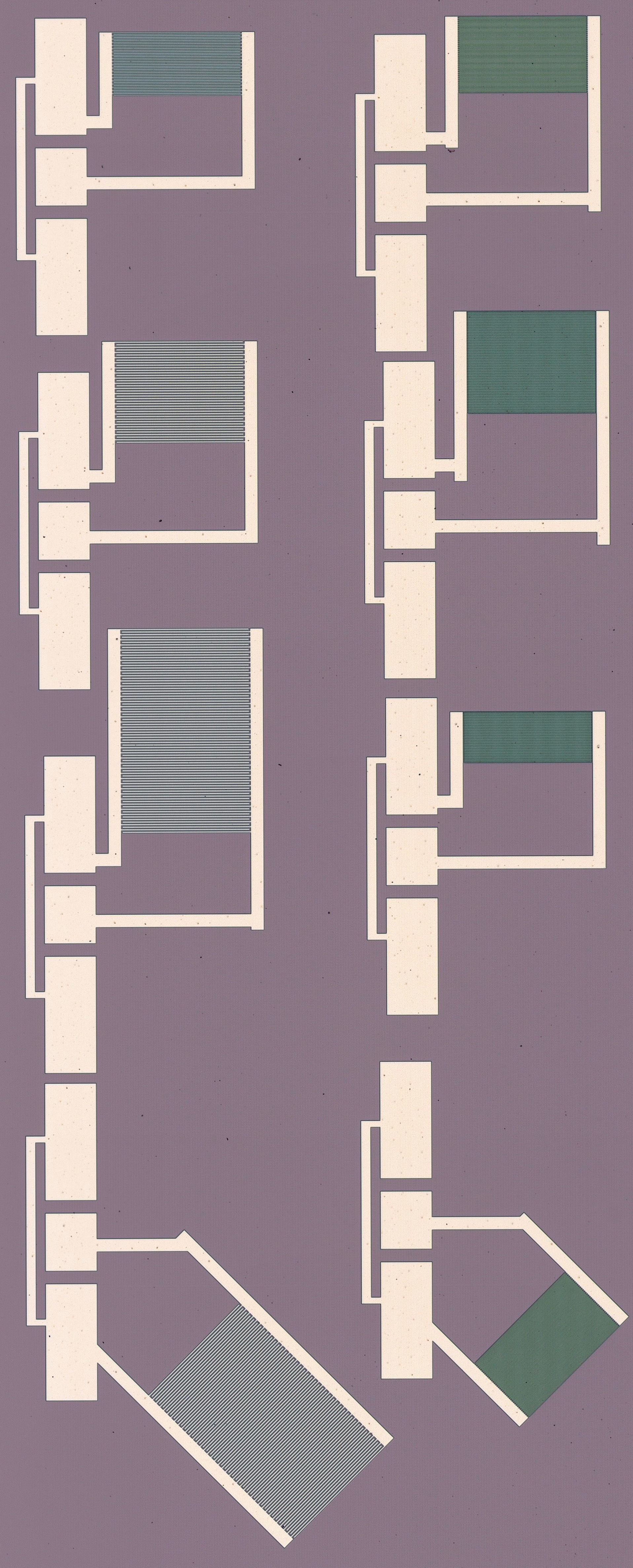}
        \label{fig:left}
    \end{subfigure}
    \hspace{0.5cm}
    \begin{subfigure}[b]{0.367\textwidth}
        \centering
        \includegraphics[width=\textwidth]{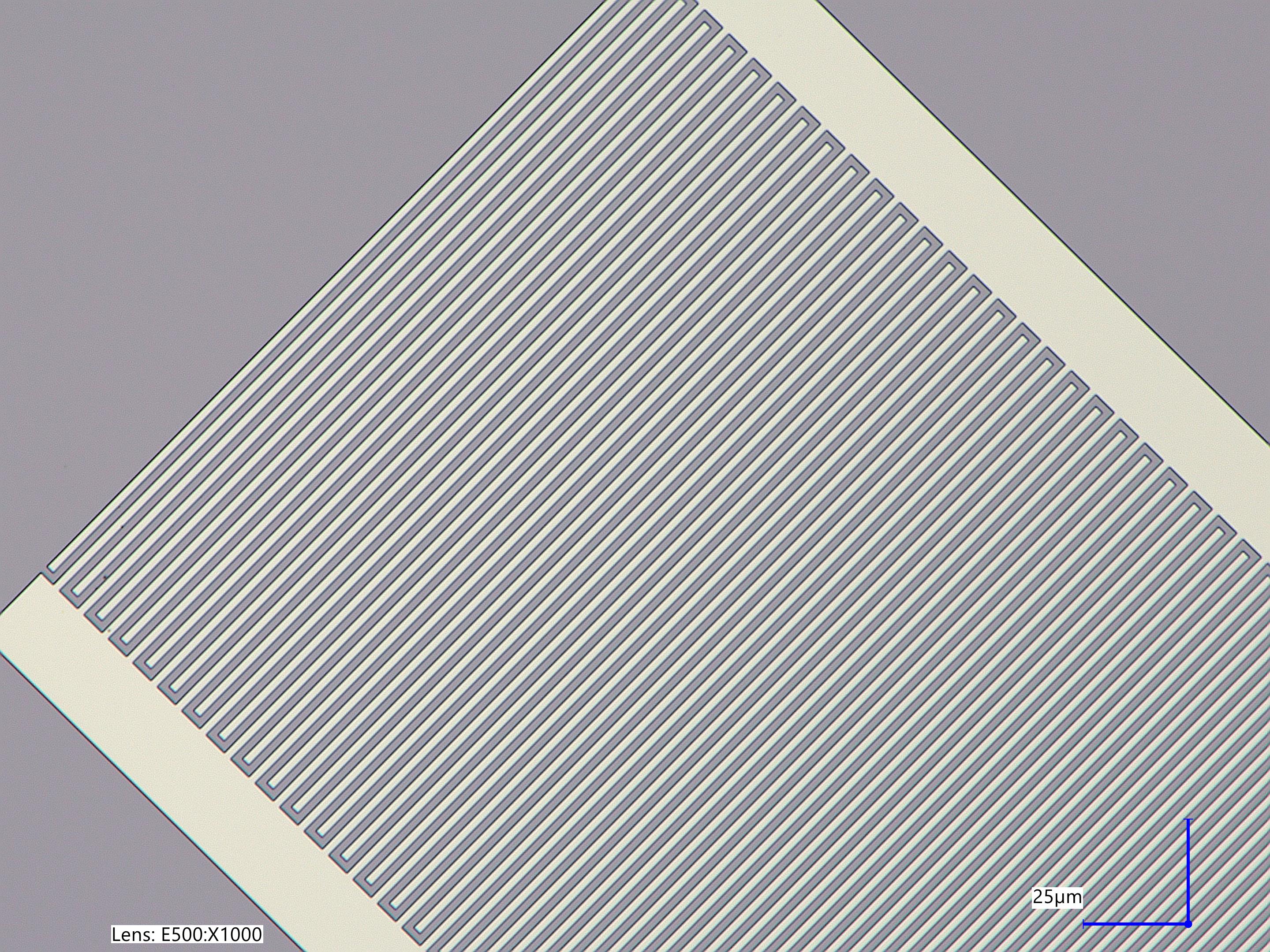}
        \label{fig:top_right}
        \vspace{1em}
        
        \includegraphics[width=\textwidth]{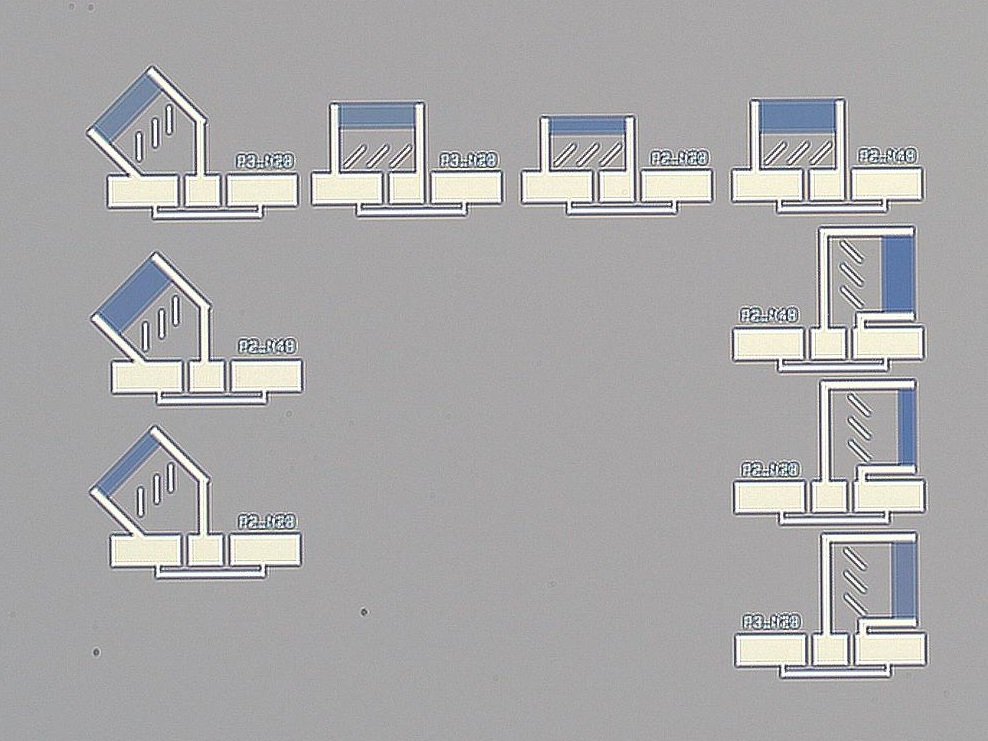}
        \label{fig:bottom_right}
    \end{subfigure}%
    }
    \caption{\textbf{IDT on SrTiO$_3$ thin film.} Microscope image of a representative IDT fabricated on a SrTiO$_3$ thin film.}
    \label{fig:IDT_STO}
\end{figure}

The fabricated IDTs (Fig.~\ref{fig:IDT_STO}) were designed with a fixed aperture width $W=~200~\mu$m and a duty cycle of 50\%.
The finger periodicity ($\Lambda$) was varied from 2~$\mu$m to 8~$\mu$m in order to span
a range of acoustic wavelengths, with a total number of finger pairs $N = 20$, 40, 
and 60 to study impedance matching for 0$^\circ$, 90$^\circ$ and 45$^\circ$ orientations. 

\subsection{Electrical characterization and impedance extraction}

All devices were characterized at a temperature of 5~K  using
an Attocube attoDRY800 system and Keysight (P5003B) vector network analyzer (VNA). 
Using the measured reflection coefficient $S_{11}(f)$, calibrated up to the device under test, 
we extract the IDT impedance as
\begin{equation}
Z(f) = Z_0 \frac{1 + S_{11}(f)}{1 - S_{11}(f)},
\end{equation}
where $Z_0 = 50~\Omega$ is the characteristic impedance of the measurement system.

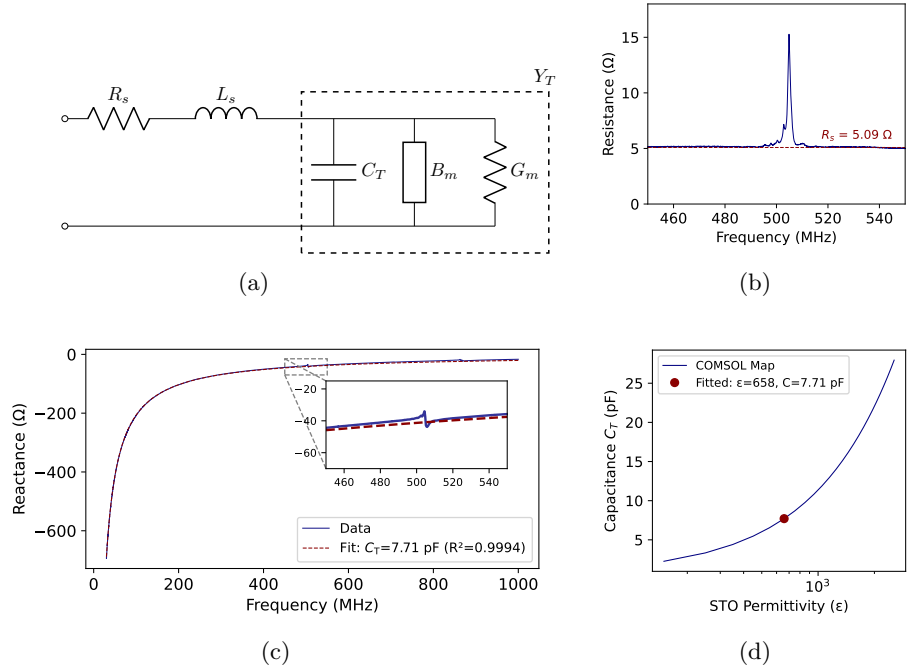
\begin{figure}[H]
    \centering

    \begin{subfigure}[b]{0.55\textwidth} \hspace*{7mm}
        \centering
        \resizebox{\linewidth}{!}{%
        \begin{circuitikz}[american]
            \draw (0,2) node[ocirc]{} to[R=$R_s$] (2,2)
                  to[L=$L_s$] (4,2) -- (8,2);
            \draw (0,0) node[ocirc]{} -- (8,0);

            \draw (5,2) to[C=$C_T$] (5,0);
            \draw (6.5,2) to[generic,l=$B_m$] (6.5,0);
            \draw (8,2) to[R=$G_m$] (8,0);

            \draw[dashed, thick] (4.4,-0.5) rectangle (9,2.5);
            \node[above right] at (8.6,2.5) {$Y_T$};
        \end{circuitikz}%
        }
        \caption{}
        \label{fig:left_top}
    \end{subfigure}
    \hfill
    \begin{subfigure}[b]{0.351\textwidth}\hspace*{-1.5mm}
        \centering
        \includegraphics[width=\linewidth]{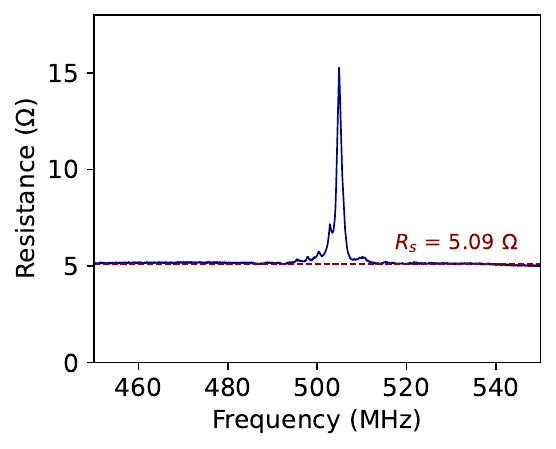}
        \caption{}
        \label{fig:right_top}
    \end{subfigure}

    \vspace{0.5cm}

    \begin{subfigure}[b]{0.598\textwidth}
        \centering
        \includegraphics[width=\linewidth]{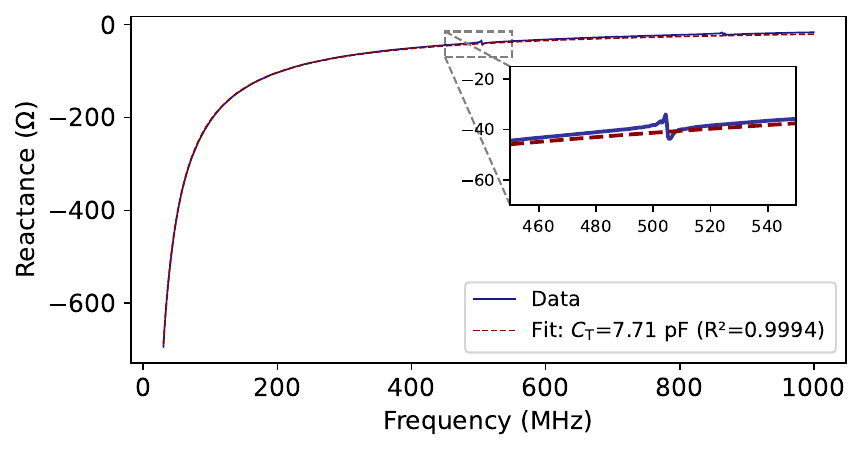}
        \caption{}
        \label{fig:left_bottom}
    \end{subfigure}
    \hfill
    \begin{subfigure}[b]{0.351\textwidth}
        \centering
        \includegraphics[width=\linewidth]{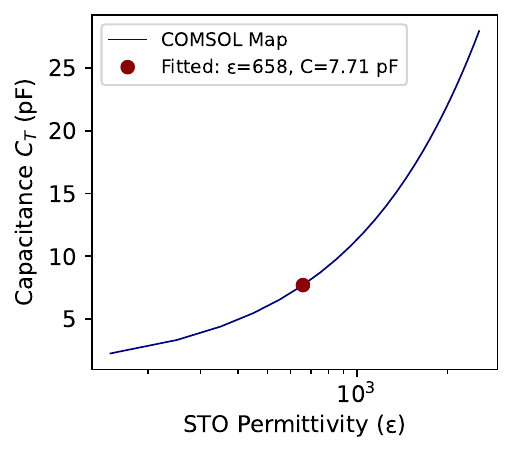}
        \caption{}
        \label{fig:right_bottom}
    \end{subfigure}

     \caption{\textbf{Mason model and permittivity extraction.} (a) Mason equivalent circuit model for the IDT electromechanical response. The typical measured resistance and wideband $\mathrm{Im}[Z]$ are shown in (b) and (c), respectively. (d) Shows the numerically simulated map from capacitance to SrTiO$_3$ permittivity.}
    \label{fig:Mason}
\end{figure}

\subsection{Mason model and electromechanical coupling}

The electromechanical response of the IDTs is better understood using Mason’s equivalent circuit model, 
in which the total electrical admittance $Y(f)$ is expressed as the sum 
of a static capacitive contribution and a frequency-dependent motional response~\cite{Mason}:
\begin{equation}
Y(f) =  i \omega C_T+ G_m(f) + i B_m(f) ,
\end{equation}
where $C_T$ is the static (geometric) capacitance of the IDT, 
and $G_m(f)$ and $B_m(f)$ are the motional conductance and susceptance, respectively.
The motional conductance $G_m(f)$ quantifies the conversion of electrical energy
into mechanical (acoustic) energy and is given by
\begin{equation}
G_m(f) = 8 K^2 C_T N f_0
\left(\frac{\sin X}{X}\right)^2
\equiv G_0 \left(\frac{\sin X}{X}\right)^2,
\qquad
X = N\pi \frac{|f - f_0|}{f_0},
\end{equation}
where $K^2$ is the electromechanical coupling coefficient, 
$f_0$ is the acoustic resonance frequency, and $G_0$ denotes the peak motional conductance. 
By causality, the motional susceptance $B_m(f)$ is related to $G_m(f)$ via 
a Hilbert transform and is given by
\begin{equation}
B_m(f) = G_0 \frac{\sin(2X) - 2X}{2X^2}.
\end{equation}
Both $G_m(f)$ and $B_m(f)$ are narrow-band features centered around $f_0$ 
with a fractional bandwidth of \(\Delta f / f_0\approx0.885 / N\). Fig.~\ref{fig:Mason}(a)
shows a circuit representation of the Mason model with addition of parasitic resistance ($R_s$) 
and inductance ($L_s$). Fig.~\ref{fig:Mason}(b,c) show a typical measured resistance $\text{Re}[Z]$, and wide-band Im$[Z]$, respectively. 

As evident from Fig.~\ref{fig:Mason} (b), the parasitic resistance contributes a constant background to the measured response. In the following, we focus on extracting the motional resistance that appears on top of this background. At resonance, the motional resistance is given by\begin{equation}
R_m (f_0)=\Re\{1/Y(f_0)\}=\frac{G_0}{G_0^2+\left(\omega_0 C_T\right)^2}.
\end{equation}
The motional resistance is governed by the electromechanical coupling coefficient $K^2$,
which itself is determined by the piezoelectric tensor ($\mathbf{e}$), 
permittivity ($\mathbf{\epsilon}$) and elasticity ($\mathbf{c}$) tensors of the SrTiO$_3$ film. In bulk acoustic wave transducers,
 $K^2$ is given by $K^2 =\frac{e_{33}^2}{c_{33} \varepsilon_{33}}$ ~\cite{BAWD}.  
However, for surface acoustic wave (SAW) transducers, $K^2$ does not have a closed-form expression and can be regarded
as a function $K^2 = f (\mathbf{e,\epsilon,c})$. Therefore, we numerically simulate the IDT electromechanical 
response using 2D-FEM in COMSOL Multiphysics to model the Rayleigh wave and extract the SrTiO$_3$ material properties. The entire number of fingers was simulated to calculate the motional resistance. 
Since the simulation is two dimensional, the shear waves (out-of-plane in the 2D simulation) is not captured.
The required permittivity extraction is extracted using the COMSOL map in Fig.~\ref{fig:Mason}(d) as discussed earlier in the methods section. The elasticity tensor is taken from literature~\cite{elastic1,elastic2}, namely, $c_{11} = 337.23$ [GPa], $c_{12} = 132.2$ [GPa], $c_{44}= 115.34$ [GPa].
In the simulation, the elasticity of SrTiO$_3$ was modeled as a "cubic" tensor, where the diagonal elements satisfy $c_{11}=c_{22}=c_{33}$, the off-diagonal elements $c_{ij}$ (for $i \neq j$ and $i,j=1,2,3$)
 are equal, and the shear components satisfy $c_{44} = c_{55} = c_{66}$. 
 This approach shows a good agreement with our acoustic dispersion results presented in the main text. The only free parameter 
is the piezoelectric tensor $\mathbf{e}$, which is adjusted to fit the amplitude of the measured motional resistance.

For Rayleigh-wave excitation, the dominant piezoelectric tensor components are $e_{33}$ and $e_{31}$,
consistent with the predominantly in-plane displacement of Ti ions in the SrTiO$_3$ thin films as discussed in section \ref{sec:tensor}.
The $e_{31}$ component is negative and can be physically understood as the transverse contraction of the unit cell accompanying expansion 
along the polar axis.
For ABO$_3$ perovskites, $|e_{31}|$ is typically at least one order of magnitude smaller than $e_{33}$~\cite{BTO}.
Density functional theory calculations for SrTiO$_3$ suggest even smaller values of $e_{31}$~\cite{DFT}.
In our analysis, we therefore assume $e_{31} = -0.1\,e_{33}$, as an upper bound estimation for $e_{31}$ (lower bound for $e_{33}$) based on related perovskite materials.
In summary, the extraction methodology for $e_{33,\mathrm{eff}}$ is as follows:\\
Analyzing wide-band data:
\[
\Im{[Z]} \xrightarrow{\text { Fitting }} C[p F] \xrightarrow{\text { COMSOL }} \epsilon_{\mathrm{STO}}
\]
Analyzing near-resonance data:
\[
\Re{[Z]}_{\text{No parasitics}} \xrightarrow{\operatorname{COMSOL}\left(\epsilon_{\mathrm{STO}}\right)} e_{33,\mathrm{eff}}
\]

\subsection{$e_{15}$ extraction based on BAW}
\begin{figure}[h] 
    \centering
    \includegraphics[width=0.5\textwidth]{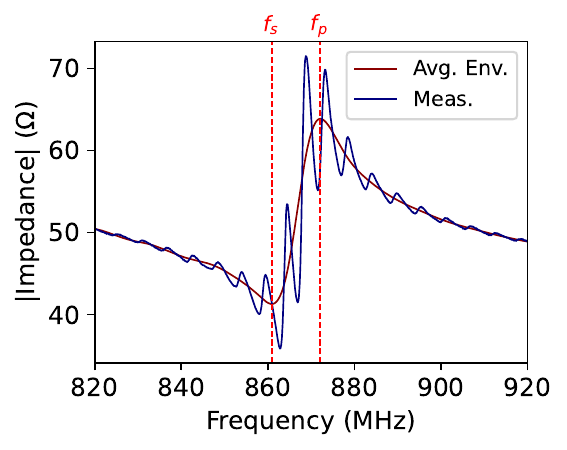}
    \caption{\textbf{$e_{15}$ extraction based on LFE.} Measured impedance for LFE IDT with $\Lambda = 8~\mu$m at 5K with $13\;V/\mu$m bias.}
    \label{fig:e15}
\end{figure}

The transducer can also be considered as a lateral field excitation (LFE) transducer with parallel electrodes configuration.
In this case, the analysis is simpler, the electromechanical coupling coefficient for thickness shear wave excitation is given by \cite{BAWD,LFE}
\begin{equation}
K_{15,\mathrm{eff}}=\frac{e_{15,\mathrm{eff}}}{\sqrt{c_{55}^{E} \varepsilon_{11}^{T}}}.
\end{equation}
Additionally, the $K_{15,\mathrm{eff}}$ can be experimentally measured using the frequency separation between the parallel and series resonance frequencies
\begin{equation}
    K_{15,\mathrm{eff} }^2=\left(\frac{\pi}{2}\right)^2 \frac{f_p - f_s}{f_p},
\end{equation}
where $f_p$ is the parallel resonance frequency and $f_s$ is the series resonance frequency shown in Fig.~\ref{fig:e15}.
The figure shows an FSR of 4.2 MHz due to the back-reflections from the Si substrate. 
In order to mitigate the uncertainty in calculating the frequency separation,
 the separation is extracted by averaging the upper and lower impedance envelopes.
 This procedure is justified by comparing the results of Mason’s model for bounded and unbounded transducers.
\subsection{Acoustic wave excitation dependency on piezoelectric tensor}\label{sec:tensor}
\begin{figure}[t]
\centering
\begin{tikzpicture}
    \node[anchor=south west, inner sep=0] (image) at (0,0) 
        {\includegraphics[width=\textwidth]{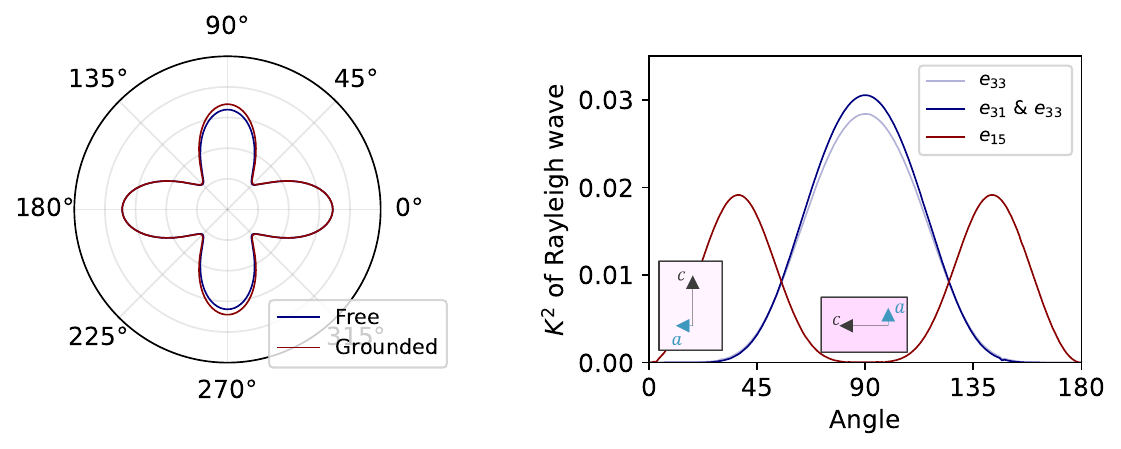}};
    \begin{scope}[x={(image.south east)}, y={(image.north west)}]
        \node at (0.02, 0.98) {(a)};
        \node at (0.52, 0.98) {(b)};
    \end{scope}
\end{tikzpicture}
\caption{\textbf{Rayleigh wave slowness curves and coupling.} (a) Slowness curves for Rayleigh wave with and without piezoelectric coupling accounting only $e_{33}$ component. (b) Rayleigh wave electromechanical coupling for different piezoelectric tensor elements.}
\label{fig:slowness}
\end{figure}

To understand the dependency of Rayleigh wave excitation on different piezoelectric tensor components, we solve the governing 
equations for acoustic wave as an eigenvalue problem to obtain the acoustic modes' profiles and their corresponding eigenfrequencies (velocities) in COMSOL Multiphysics, namely \cite{Zhang2022-bc},
\begin{equation}
\begin{aligned}
\nabla_s\left(\mathbf{c}\left[\nabla_s\right]^T \mathbf{u}+\mathbf{e}^T[\nabla]^T \phi\right)+\rho \omega^2 \mathbf{u} & =0, \\
\nabla\left(\mathbf{e}\left[\nabla_s\right]^T \mathbf{u}-\varepsilon[\nabla]^T \phi\right) & =0 .
\end{aligned}
\end{equation}
where $\mathbf{u}$ is the displacement vector, $\phi$ is the electric potential, $\mathbf{c}$ is the elasticity tensor, $\mathbf{e}$ is the piezoelectric stress-tensor, $\boldsymbol{\varepsilon}$ is the permittivity tensor and $\rho$ is the density.
For a given acoustic mode, the magnitude of the gap between dispersion curves calculated with and without piezoelectric
 coupling provides a direct measure of how efficiently that mode can be electrically excited.
Piezoelectric coupling modifies the acoustic phase velocity;
 by reciprocity, this implies that an electric field applied along the corresponding crystallographic direction can efficiently drive the same acoustic mode.
Guided by this principle, we perform single-domain
 simulations both including piezoelectric coupling,
  implemented via free-floating electric potential boundary conditions (BCs), and excluding piezoelectric coupling, enforced through grounded potential BCs.
   This procedure enables the systematic construction of slowness curves (inverse phase velocity) 
   as a function of propagation direction and allows the contribution of individual piezoelectric
    tensor components to be isolated.
Fig.~\ref{fig:slowness}a shows the slowness curves of the Rayleigh wave under free and grounded potential BCs, considering only the $e_{33}$ component.
  It follows that the electromechanical coupling can be calculated from the slowness curves as \cite{Zhang2022-bc}
\begin{equation}
    K^2=2 \times \left(1-\frac{v_{\mathrm{grounded}}}{v_{\mathrm{free}}}\right) \times 100\%.
\end{equation}

\begin{figure}
    \centering
    \includegraphics[width=0.7\textwidth]{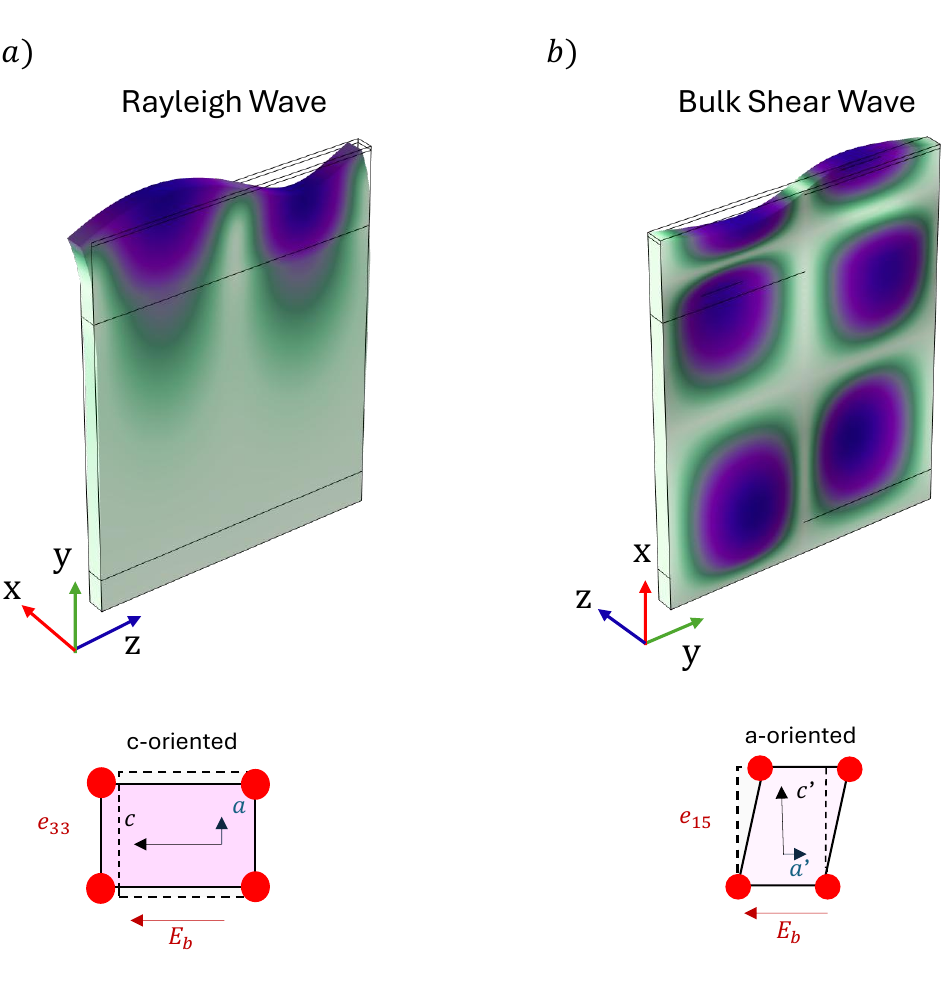}
    \caption{\textbf{Simulated acoustic mode profiles.} Simulated mode profiles with its corresponding driving domain for (a) Rayleigh wave and (b) Bulk shear wave.}
    \label{fig:mode_profiles}
\end{figure}

Fig.~\ref{fig:slowness}(b) shows the dependency of the Rayleigh wave excitation on the piezoelectric coefficient $e_{33}$. The wave can be excited efficiently in the c-oriented domain but not in the a-oriented domain. 
It can be seen that the $e_{31}$ component slightly enhances the excitation, while the $e_{15}$ component doesn't contribute to the Rayleigh wave excitation with 0 and 90 degrees IDTs.  
Similar interpretations can be drawn by merely looking at the rotated piezoelectric tensor. As a reminder, the piezoelectric tensor for tetragonal c-axis (Ti atom is out of the wafer's plane) SrTiO$_3$ is given by
\begin{equation}
[e]=\begin{bmatrix} 
0 & 0 & 0 & 0 & e_{15} & 0 \\
0 & 0 & 0 & e_{15} & 0 & 0 \\
e_{31} & e_{31} & e_{33} & 0 & 0 & 0
\end{bmatrix}.
\end{equation}
For a c-axis film, the crystallographic axes $(1,2,3)$ are parallel to the electric-field coordinate system, corresponding to ($1 = x$, $2 = y$, and $3 = z$).
However, our grown SrTiO$_3$ is a multidomain film with Ti atoms oriented in-plane (a-axis) either with 0$^\circ$ (regarded a-oriented) as 
\begin{equation}
[e]_a^0=\begin{array}{cccccc}
S_1 \quad S_2 \quad S_3 \quad S_4 \quad S_5 \quad  S_6\\

\left[\begin{array}{cccccc}
0 & 0 & 0 & 0 & 0 & \color{red}e_{15} \\
e_{31} & e_{33} & e_{31} & 0 & 0 & 0 \\
0 &0& 0 & e_{15} & 0 & 0\\
\end{array}\right] \\
\end{array}
 \begin{gathered}
\\
\color{blue}E_x \\
E_y \\
E_z
\end{gathered}
\end{equation}
or with 90$^\circ$ (regarded c-oriented) as
\begin{equation}
[e]_c^{90}=
\begin{array}{cccccc}
S_1 \quad S_2 \quad S_3 \quad S_4 \quad S_5 \quad  S_6\\

\left[\begin{array}{cccccc}
\color{red}e_{33} & \color{red}e_{31} & \color{red}e_{31} & 0 & 0 & 0 \\
0 & 0 & 0 & 0 & 0 & e_{15} \\
0 &0& 0 & 0 & e_{15} & 0\\
\end{array}\right] \\
\end{array}
 \begin{gathered}
\\
\color{blue}E_x \\
E_y \\
E_z
\end{gathered}
\end{equation}
For a dominantly x-oriented electric field across $[e]_a^0$
the $e_{15}$ now is at 1,6 position which couples x-field to shear x-y stress, consistent with the mode profile shown in Fig.~\ref{fig:mode_profiles}a.
Similarly, for x-oriented electric field across $[e]_c^{90}$, the $e_{33}$ now is at 1,1 position which couples x-field to x -longitudinal stress,  
and $e_{31}$ at the 1,3 position which couples x-field to z- transverse stresses. This matches the elliptical motion that builds the Rayleigh wave.
\subsection{Single domain piezoelectric coefficient of twin-domain structure}
Starting from the electromechanical coupling coefficient definition \cite{Berlincourt19643P}
\begin{equation}
K=\frac{U_m}{\sqrt{U_e U_d}},
\end{equation}
where $U_m$ is the mutual energy, $U_e$ is the elastic energy and $U_d$ is the dielectric energy.
For a twin-domain SrTiO$_3$ thin film with $a$- and $c$-oriented domains, the energies for an electric field oriented along the crystallographic $c$ axis are given by
\begin{equation}
U_m=\frac{1}{2}\left[\int_{\Omega_a}\left(\frac{e_{33}^{(a)}}{c_{33}^{E(a)}}\right) E_3 T_3 d V+\int_{\Omega_c}\left(\frac{e_{33}^{(c)}}{c_{33}^{E(c)}}\right) E_3 T_3 d V\right],
\end{equation}
\begin{align}
U_e &= \frac{1}{2} \int_{\Omega} \frac{1}{c_{33}^{E}} T_3^2 d V, \\
U_d &= \frac{1}{2} \int_{\Omega} \varepsilon_{33}^{T} E_3^2 d V.
\end{align}
Here, $E_3$ and $T_3$ are the electric field and the stress in the c-axis, respectively.
For our twin boundary, the volumetric fraction between $\mathbf{(c)}$-and $\mathbf{(a)}$-oriented domains is 50\% in the case of 0 and 90 degree IDTs
\begin{equation}
\operatorname{Vol}\left(\Omega_a\right)=\operatorname{Vol}\left(\Omega_c\right)=\frac{1}{2} \operatorname{Vol}(\Omega).
\end{equation}
Since a-oriented domains do not excite Rayleigh waves ($e_{33}^{(a)} = 0$),
only the $c$-domain volume contributes to the mutual energy, while the elastic and dielectric energies are calculated over both c and a volume/domain.
As a result, the electromechanical coupling coefficient that was utilized in the acousto-optic modulator was reduced by a factor of two relative to a single-domain $c$-oriented film
\begin{equation}
K_{33,\mathrm{eff}}= \frac{1}{2} K_{33}^{(c)}.
\end{equation}
Equivalently, the effective piezoelectric coefficient satisfies $e_{33,\mathrm{eff}} = e_{33}^{(c)}/2$.

Similarly, for the $e_{15}$ with a 45$^\circ$ IDT, the mutual energy is given by
\begin{equation}
U_m=\frac{1}{2}\left[\int_{\Omega_x}\left(\frac{e_{15}^{(x)}}{c_{55}^{E(x)}}\right) \cos(45^\circ)^2 E_1 T_5 d V+\int_{\Omega_y}\left(\frac{e_{15}^{(y)}}{c_{55}^{E(y)}}\right) \cos(45^\circ)^2 E_1 T_5 d V\right].  
\end{equation}
where the x, and y are the in-plane orientation 0$^\circ$,and 90$^\circ$. The total energies are,
 \begin{align}
U_e &= \frac{1}{2} \int_{\Omega} \frac{1}{c_{55}^{E}} T_5^2 d V, \\
U_d &= \frac{1}{2} \int_{\Omega} \varepsilon_{11}^{T} E_1^2 d V.
\end{align}
Since both directions exhibit identical response ($e_{15}^{(x)} =e_{15}^{(y)} = e_{15}^{(a)}$), while the volumetric fraction of the domains is still 50\%,
\begin{equation}
K_{15,\mathrm{eff}}= \frac{1}{2} K_{15}^{(a)}.
\end{equation}
Equivalently, the effective piezoelectric coefficient satisfies $e_{15,\mathrm{eff}} = e_{15}^{(a)}/2$.

\subsection{Acoustic propagation losses}
A propagating acoustic mode in a multilayer structure can be described by a
complex wavevector
\begin{equation}
\tilde{k} = \beta + i\alpha ,
\end{equation}
where $\beta$ is the real propagation constant and $\alpha>0$ quantifies spatial
attenuation. The displacement field therefore decays as
\begin{equation}
u(x) = u_0 e^{i\beta x} e^{-\alpha x}.
\end{equation}
Since the strain field $\varepsilon(x)$ is proportional to the spatial derivative
of the displacement, $\varepsilon \sim \partial u/\partial x$, its amplitude
inherits the same exponential decay,
\begin{equation}
\varepsilon(x) \propto e^{-\alpha x}.
\end{equation}
We thus define the \emph{strain amplitude decay length} as
\begin{equation}
L_A \equiv \frac{1}{\alpha},
\end{equation}
corresponding to the propagation distance over which the strain amplitude decays
to $1/e$ of its initial value. The quality factor $Q$ of the propagating mode can be expressed directly in terms
of the real and imaginary parts of the wavevector as
\begin{equation}
Q = \frac{\beta}{2\alpha},
\end{equation}
\begin{figure}[!b]
    \centering
    \includegraphics[width=0.5\textwidth]{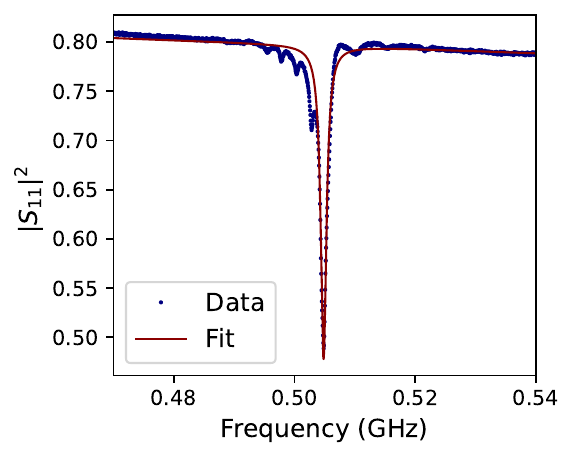}
    \caption{\textbf{Quality factor of the acoustic resonance.} Measured acoustic resonance at $3~V/\mu$m bias. Solid line represents a Lorentzian fit.}
    \label{fig:losses}
\end{figure}
which reflects the ratio between phase accumulation and amplitude loss per
propagation cycle. Solving for $\alpha$ and substituting into the definition of
$L_A$ yields
\begin{equation}
L_A = \frac{2Q}{\beta}=\frac{Q\,\lambda}{\pi}.
\end{equation}
In our device shown in Fig.~\ref{fig:losses} ($f_0=504~\mathrm{MHz}$, $\gamma/2\pi=1.136~\mathrm{MHz}$), the loaded
quality factor is $Q_L=f_0/(\gamma/2\pi)\approx 444$. With an external coupling fraction
$\gamma_{\mathrm{ext}}/\gamma=0.1447$, we obtain $Q_i=Q_L/(1-\gamma_{\mathrm{ext}}/\gamma)\approx 519$.
For an acoustic wavelength $\lambda=8~\mu\mathrm{m}$, the intrinsic strain decay length is
\begin{equation}
L_{A,i}=\frac{Q_i\lambda}{\pi}\approx \frac{519\times 8~\mu\mathrm{m}}{\pi}\approx 1.32~\mathrm{mm}.
\end{equation}

\section{Acousto-optic modulation}
\subsection{Calculating the $V_\pi L$}
\begin{figure}[h]
    \centering
    \begin{subfigure}[b]{0.44\textwidth}
        \centering
        \begin{tikzpicture}
            \node[anchor=south west, inner sep=0] (image) at (0,0)
                {\includegraphics[width=\textwidth]{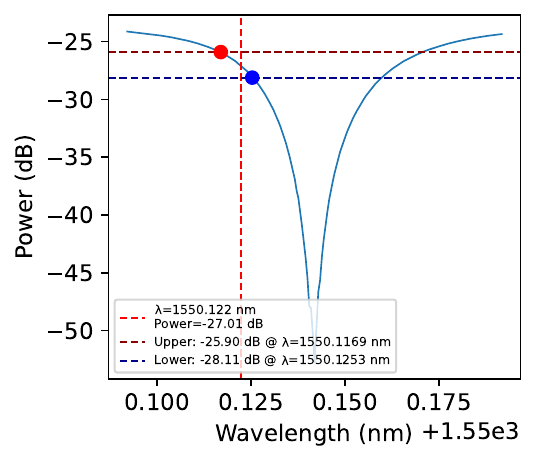}};
            \begin{scope}[x={(image.south east)}, y={(image.north west)}]
                \node at (0.05, 0.95) {(a)};
            \end{scope}
        \end{tikzpicture}
        \label{fig:AOM_setup_a}
    \end{subfigure}%
    \hspace{0.5cm}
    \begin{subfigure}[b]{0.51\textwidth}
        \centering
        \raisebox{1.5cm}{%
        \begin{tikzpicture}
            \node[anchor=north west, inner sep=1] (image) at (0,0)
                {\includegraphics[width=\textwidth]{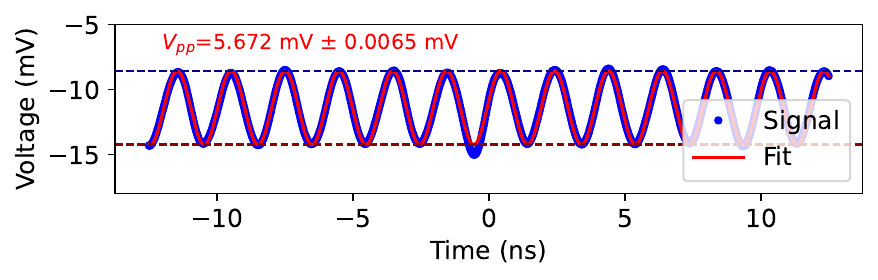}};
            \begin{scope}[x={(image.north east)}, y={(image.north west)}]
                \node at (0.01, 1) {(b)};
            \end{scope}
        \end{tikzpicture}}
        \label{fig:AOM_setup_b}
    \end{subfigure}
    \caption{\textbf{Acousto-optic modulation measurement.} (a) Racetrack response showing the 8.4 pm wavelength shift. (b) Averaged oscilloscope output voltage.}
    \label{fig:AOM_setup}
\end{figure}

To accurately calculate the $V_\pi L$ of the acousto-optic modulator, we measure the electrical output coming from the photodetector using an oscilloscope as shown in Fig.~\ref{fig:AOM_setup}.
We modulate the ring resonator at the 3-dB point to maximize the modulation depth while driving the RF at the Rayleigh wave resonance frequency with 1 dBm RF power.
Using the ring transfer function, we can relate the photodetector output peak to peak voltage ($V_{pp}$) to the wavelength shift ($\Delta \lambda$) caused by the $\Delta n_{\mathrm{eff}}$.

Since the photodetector voltage is linear with the optical power, a change of 2.21~dB in the output voltage corresponds to an identical 2.21~dB change in the optical power. This power variation corresponds to a wavelength shift of $\Delta \lambda = 8.4~\mathrm{pm}$.

The power and thus the driving voltage delivered to the device can be calculated from the $S_{11}$ measurement as
\begin{equation}
P_{\mathrm{device}} = P_{\mathrm{input}} (1 - |S_{11}|^2),
\end{equation}
where $P_{\mathrm{input}}$ is the input RF power of the VNA. The peak voltage is defined as $V_{p} = \sqrt{2\;Z_L\;P_{\mathrm{device}}}$ where $Z_L$ is the IDT impedance.
The calculated peak voltage is $V_{p} = 0.235\;V$ for 1 dBm input power. The $V_\pi L$ can be calculated as
\begin{equation}
V_\pi L = \frac{FSR \; V_{p}\; L}{\Delta \lambda},
\end{equation}
where $FSR$ denotes the free spectral range of the ring resonator,
 $FSR = 0.781~\mathrm{nm}$, and $L_c$ is the interaction length between the optical mode and the acoustic wave.
  Using the above equation, we extract $V_{\pi}L = 0.874 \pm 0.084~\mathrm{V\cdot cm}$ for an interaction length
   of $L = 2\times 200~\mu\mathrm{m}$. 

Because the acoustic wave can propagate reaching the second arm of the racetrack resonator with negligible decay, the effective interaction length is doubled.
Assuming negligible acoustic dispersion in the CSAR layer, the acoustic wavelength ($8~\mu$m) is an integer divisor of the $400~\mu$m separation between the two arms.
This configuration thus realizes a push-push modulation scheme, in which the refractive index modulation in both arms is matched in amplitude and phase.

\subsection{Photoelasticity}
The aforementioned wavelength shift, $\Delta\lambda$, arises from a change in the effective refractive 
index $n_{\mathrm{eff}}$ of the ring resonator. 
The relationship between $\Delta \lambda $ and $\delta n_{\mathrm{eff}}$ can be obtained by writing the resonance phase condition for the $m^{\text{th}}$ azimuthal mode in the resonator
\begin{equation}
\beta^{(0)}\left(\omega_0\right) L=2 \pi m,
\end{equation}
while the perturbed state satisfies
\begin{equation}
\beta^{(1)}\left(\omega_1\right) L_c+\beta^{(0)}\left(\omega_1\right)\left(L-L_c\right)=2 \pi m,
\end{equation}
where $L$ is the circumference of the racetrack, given by its radius \(R\) and straight length \(L_c/2\) (i.e. L= \(2\pi R +L_c\)), and \(\beta^{(i)}(\omega_j)\) is the propagation
 constant of the waveguide mode (\(i=0\) non-perturbed, 
 \(i=1\) perturbed) evaluated at a specific frequency.
  Subtracting both equations we get 
\begin{equation}
  \begin{aligned}
 \left(\beta^{(1)}\left(\omega_1\right)-\beta^{(0)}\left(\omega_1\right)\right) L_c +\left(\beta^{(0)}\left(\omega_1\right)-\beta^{(0)}\left(\omega_0\right)\right) L=&\;0 \\
 \left(\frac{\omega_0}{c} \delta n_{e f f}\right) L_c + \left(\frac{n_g}{c} \delta \omega\right) L = & \;0,
\end{aligned}
\end{equation}
we obtain
\begin{equation}
\delta n_{\mathrm{eff}} = -n_g\frac{\delta \omega}{\omega}\frac{L}{L_c},
\end{equation}
where \(n_g\) is the group index and \(\omega\) is the mode frequency.
 The term \(\delta\omega/\omega\) can be cast into a wavelength form with a change of variables 
 (\(\delta\omega/\omega=-\delta\lambda/\lambda\)), while
\(n_g\) can be inferred from the \(\mathrm{FSR}\) via \(n_g=\lambda^2/(\mathrm{FSR}\times L)\). 
resulting in the final expression for the measured \(\delta n_{eff}\) as
\begin{equation}
\delta n_{\mathrm{eff}} = \frac{\lambda\;\delta\lambda}{FSR\times L_c}.
\end{equation}
Hence, the total effective refractive index modulation can be directly obtained from the measurement.

\(\delta n_{\mathrm{eff}}\) can also be inferred from perturbation theory \cite{Shao}
\begin{equation}
\delta n_{\mathrm{eff}} = \frac{n_{\mathrm{eff}}}{2} \frac{\left\langle E\right| \Delta \varepsilon\left|E\right\rangle}{\left\langle E\right| \varepsilon\left|E\right\rangle},
\label{eq:freq-shift}
\end{equation}
where \(\Delta\varepsilon\) is the permittivity perturbation and \(\left|E\right\rangle\)
is the unperturbed field.
The effective refractive index modulation \(\delta n_{\mathrm{eff}}\) consists of
 contributions arising from moving boundaries \((\delta n_{\mathrm{eff,MB}})\) 
 and from the photoelastic effect \((\delta n_{\mathrm{eff,PE}})\).
  The photoelastic contribution is further decomposed into the contributions originating
   from the CSAR waveguide \((\delta n_{\mathrm{eff,CSAR}})\),
    the \(\mathrm{SiO}_2\) substrate \((\delta n_{\mathrm{eff,SiO_2}})\), and the SrTiO$_3$ layer \((\delta n_{\mathrm{eff,STO}})\).

To isolate \(\delta n_{\mathrm{eff,STO}}\), all other contributions are calculated independently using finite‑element simulations. 
A surface acoustic wave is simulated in the cross‑section of the ring resonator to obtain the strain and displacement fields at the waveguide, 
while a separate electromagnetic eigenmode simulation provides the unperturbed optical field.
 The individual contributions are subsequently evaluated by post‑processing the simulated fields. 
In this simulation, the extracted effective piezoelectric and dielectric constant of the experiment are used at $3~V/\mu m$  ($e_{33,\mathrm{eff}} = 5.52\;\mathrm{C/m}^{2}$  \& $\epsilon = 658$).
\subsubsection{Moving boundary contribution}
The perturbation term in Eq.~\eqref{eq:freq-shift} arising from domains with shifting material boundaries was first derived in Ref.~\cite{Johnson_PRE_2002}.
 For an interface between two materials with permittivities \(\varepsilon_1\) and \(\varepsilon_2\), displaced by a mechanical displacement field \(\mathbf{u}\),
  the corresponding moving‑boundary contribution to the permittivity perturbation can be expressed as
\begin{equation}
\left\langle E\right| \Delta \varepsilon\left|E\right\rangle=\int d s\  \hat{n}\cdot\vec{u}\left[ (\varepsilon_{1}-\varepsilon_{2})\left|E_{\|}\right|^2-(\varepsilon_{1}^{-1}-\varepsilon_{2}^{-1})\left|D_{\perp}\right|^2\right],
\end{equation}
where \(\hat{n}\) denotes the unit normal to the material interface, \(E_{\parallel}\) is the component of the electric field parallel to the interface, and \(D_{\perp}\) is the component of the electric displacement field normal to the interface. 
The surface integral is evaluated over the material boundary stated as $ds$.
For the waveguide considered here, the total moving‑boundary contribution is obtained by summing the contributions from all material interfaces in the waveguide cross‑section.
 
\subsubsection{Photoelastic contribution}
The photoelastic contribution \(\delta n_{\mathrm{eff,PE},\mathrm{M}}\) of each material (M) besides SrTiO$_3$, namely CSAR and SiO$_2$ is given by
\begin{equation}
\delta n_{\mathrm{eff}, \mathrm{PE}, \mathrm{M}}
=
\frac{\epsilon_0 n_{\mathrm{M}}^4 n_{\mathrm{eff}}}{2}
\frac{\int dr
\left(
\begin{array}{lll}
E_x^{*} & E_y^* & E_z^*
\end{array}
\right)
\left(
\begin{array}{lll}
d B_1 & d B_6 & d B_5 \\
d B_6 & d B_2 & d B_4 \\
d B_5 & d B_4 & d B_3
\end{array}
\right)
\left(
\begin{array}{l}
E_x \\
E_y \\
E_z
\end{array}
\right)}
{\int E_{\;}^{\ast} \epsilon E\, dr},
\end{equation}
where \(dB_{ij} = \Delta(1/\varepsilon_{ij})\) denotes the change in the optical impermeability tensor induced by the photoelastic effect, written here in Voigt notation. 
 We consider that for all materials in the waveguide, the photoelastic effect equation can be written as:
\begin{equation}
\left(
\begin{array}{c}
dB_1 \\
dB_2 \\
dB_3 \\
dB_4 \\
dB_5 \\
dB_6
\end{array}
\right)
=
\left(
\begin{array}{cccccc}
p_{11} & p_{12} & p_{12} & 0 & 0 & 0 \\
p_{12} & p_{11} & p_{12} & 0 & 0 & 0 \\
p_{12} & p_{12} & p_{11} & 0 & 0 & 0 \\
0 & 0 & 0 & p_{44} & 0 & 0 \\
0 & 0 & 0 & 0 & p_{44} & 0 \\
0 & 0 & 0 & 0 & 0 & p_{44}
\end{array}
\right)
\left(
\begin{array}{c}
s_1 \\
s_2 \\
s_3 \\
s_4 \\
s_5 \\
s_6
\end{array}
\right),
\end{equation}
where \(s\) is the strain in the material in the Voigt notation. For the wave considered here, the only non‑zero strain components are
\(s_{11}\,(s_1)\), \(s_{33}\,(s_3)\), and \(s_{13}\,(s_5)\), resulting in the
following non‑zero components of the impermeability tensor
\begin{equation}
\begin{aligned}
dB_1 &= p_{11}s_1 + p_{13}s_3, \\
dB_3 &= p_{31}s_1 + p_{33}s_3, \\
dB_5 &= p_{44}s_5.
\end{aligned}
\end{equation}
With this, we rewrite the photoelastic contribution:
\begin{equation}
\delta n_{\mathrm{eff}, \mathrm{PE}, \mathrm{M}}=-\frac{\epsilon_0 n_{\mathrm{M}}^4 n_{\mathrm{eff}}}{2} \frac{\int d r \ [ E_x^*dB_1E_x + E_z^*dB_3E_z + dB_5(E_x^*E_z + E_xE_z^*)]}{\int E^* \epsilon E d r},
\end{equation}
which gives us the photoelastic contributions of \(\mathrm{CSAR}\) and \(\mathrm{SiO}_2\). 
In the absence of reported photoelastic coefficients for CSAR 62,
 we approximate its stress–optic response using those of 
 PMMA\cite{Rosello-Mecho:20}, a closely related polymeric e-beam resist, 
 noting that the opto-acoustic coupling is dominated by the waveguide
  core and that the resist contributes only through a weak evanescent 
  optical overlap ($< 28\%$).
\subsubsection{Effective photoelastic coefficient of SrTiO$_3$}
As the individual tensor components of the photoelastic response of
SrTiO$_3$ cannot be independently resolved, we introduce an effective
photoelastic coefficient \(p_{\mathrm{eff}}\). By setting all non‑zero elements
of the photoelastic tensor equal to \(p_{\mathrm{eff}}\), the photoelastic
contribution can be expressed as
\begin{equation}
    \begin{aligned}
\delta n_{\mathrm{eff}, \mathrm{PE}, \mathrm{STO}} &=\frac{\epsilon_0 n_{\mathrm{STO}}^4 n_{\mathrm{eff}} }{2} \frac{\int d r \ [ (s_1+s_3)(|E_x|^2 +|E_z|^2) + 2s_5\mathrm{Re}(E_x^*E_z)]}{\int E^* \epsilon E d r} p_{\mathrm{eff}}, \\
&= \mathcal{L}\;p_{\mathrm{eff}}.
\end{aligned}
\end{equation}
The constant \(\mathcal{L}\) can thus be extracted by post‑processing the
simulated strain and optical mode profiles, and subsequently used to determine
the effective photoelastic coefficient \(p_{\mathrm{eff}}\) as:
\begin{equation}
p_{\mathrm{eff}} = \frac{1}{\mathcal{L}}(\delta n_{\mathrm{eff}} - \delta n_{\mathrm{eff}, \mathrm{MB}} - \delta n_{\mathrm{eff}, \mathrm{PE}, \mathrm{CSAR}}- \delta n_{\mathrm{eff}, \mathrm{PE}, \mathrm{SiO}_2}).
\end{equation}
Table~\ref{tab:neff} summarizes the analysis parameters and results. 
Finally, we benchmark the extracted effective photoelastic coefficient
 of SrTiO$_3$ against that of BaTiO$_3$, using identical simulated strain
  and optical field distributions. 
The photoelastic tensor elements for BaTiO$_3$ are taken from
 literature~\cite{BTO}
 \begin{equation}
p_{\mathrm{eff, BTO}}=\frac{\int d r\left[E_x^* d B_{1, B T O} E_x+E_y^* d B_{2, B T O} E_y+E_z^* d B_{3, B T O} E_z+d B_{5, B T O}\left(E_x^* E_z+E_x E_z^*\right)\right]}{\int d r\left[\left(s_1+s_3\right)\left(\left|E_x\right|^2+\left|E_y\right|^2+\left|E_z\right|^2\right)+2 s_5 \operatorname{Re}\left(E_x^* E_z\right)\right]},
\end{equation}
yielding \(p_{\mathrm{eff,BTO}}=0.61\).
This direct comparison highlights the relative strength of the
 photoelastic response in SrTiO$_3$.

\begin{table}[!h]
\centering
\caption{Key parameters and extracted quantities used in this work.}
\label{tab:neff}
\begin{tabular}{lc}
\hline
Parameter & Value \\
\hline
$R$ & $200\,\mu\mathrm{m}$ \\
$L_c$ & $400\,\mu\mathrm{m}$ \\
$\lambda$ & $1550.1\,\mathrm{nm}$ \\
FSR & $0.781 \,\mathrm{nm}$ \\ 
$n_g$ & $1.857$ \\
$\delta n_{\mathrm{eff}}$ & $2.0840 \times 10^{-5}$ \\
$\delta n_{\mathrm{eff,MB}}$ & $1.263 \times 10^{-6}$ \\
$\delta n_{\mathrm{eff,PE(CSAR+SiO_2)}}$ & $4.754 \times 10^{-6}$ \\
$\mathcal{L}$ & $2.635 \times 10^{-5}$ \\
$p_{\mathrm{eff,STO}}$ & $0.5625$ \\
\hline
\end{tabular}
\end{table}

\subsection{Optical response with increased RF power}
\begin{figure}[!b]
    \centering
    \begin{subfigure}[b]{0.48\textwidth}
        \centering
        \includegraphics[width=\textwidth]{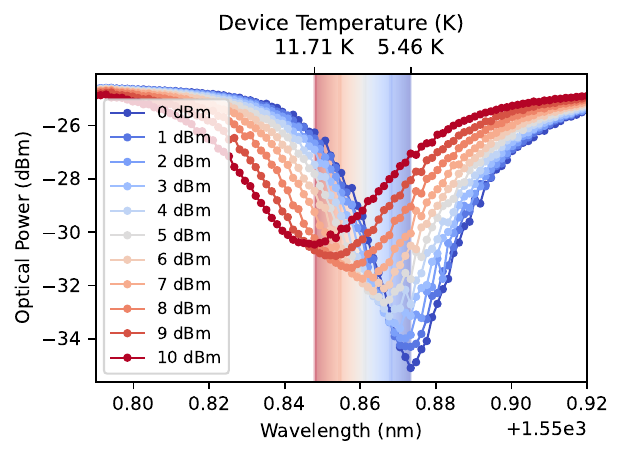}
        \caption{ \;}
        \label{fig:AOM_power_5K}
    \end{subfigure}
    \hfill
    \begin{subfigure}[b]{0.41\textwidth}
        \centering
        \includegraphics[width=\textwidth]{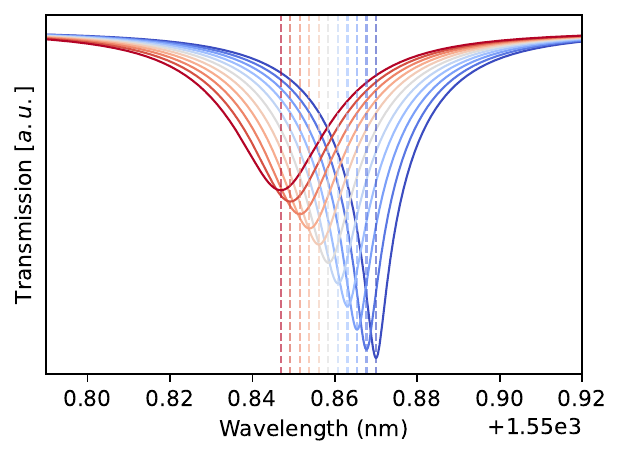}\vspace*{0.5mm}
        \caption{\;}
        \label{fig:AOM_power_RT}
    \end{subfigure}
    \caption{\textbf{Optical response vs.\ RF power.} (a) Measured optical transmission spectra of the ring resonator for different RF powers applied to the IDT. (b) Simulated optical transmission.}
    \label{fig:AOM_power}
\end{figure}
Fig.~\ref{fig:AOM_power}a shows the optical transmission as a function of the applied RF power. 
In the unresolved-sideband regime $\Omega \ll \kappa$,
 the optical cavity responds adiabatically to the mechanical motion,
  such that the optomechanical interaction induces resonance-frequency jitter 
  of the cavity. The measured transmission, which represents a time 
  average over this frequency jitter, therefore exhibits an apparent broadening
   relative to the intrinsic cavity linewidth. Additionally, we observed that
    increasing the RF power leads to a slight increase in the sample temperature, accompanied by
    a blue shift of the optical resonance. Using temperature-dependent measurements of the ring response, we correlated the observed thermal shift with the actual device temperature, estimating a temperature increase of 6.26 K at 10 dBm RF power. 

The broadening of the measured optical response under high RF power can be numerically modeled assuming 
a weak optical input using the equation of the intracavity field amplitude $a(t)$ \cite{Shao}
\begin{equation}
\dot a(t)
=
\left(i\Delta-\frac{\kappa}{2}\right)a(t)
- i\,\delta\omega \cos(\Omega t)\,a(t)
+ \sqrt{\kappa_e}\,a_{\mathrm{in}},
\label{eq:eom}
\end{equation}
where $\Delta=\omega_L-\omega_c$ is the laser--cavity detuning, $\kappa=\kappa_e+\kappa_i$ is the total optical decay rate, and $\delta\omega$ denotes the amplitude of the acoustically induced modulation of the cavity resonance frequency.
For a periodic modulation, the steady-state solution is expanded as
\begin{equation}
a(t)=\sum_n a_n e^{-i n\Omega t},
\end{equation}
yielding the coupled equations
\begin{equation}
\left(\frac{\kappa}{2}-i(\Delta+n\Omega)\right)a_n
+i\frac{\delta\omega}{2}\left(a_{n-1}+a_{n+1}\right)
=\sqrt{\kappa_e}\,a_{\mathrm{in}}\,\delta_{n0}.
\label{eq:floquet}
\end{equation}
The through-port output field is
\begin{equation}
a_{\mathrm{out}}(t)=a_{\mathrm{in}}-\sqrt{\kappa_e}\,a(t),
\end{equation}
because the optical transmission is recorded using a low-bandwidth optical power meter,
whose response time is much longer than the modulation period $2\pi/\Omega$, the measured
signal corresponds to the time-averaged transmission
\begin{equation}
T_{\mathrm{DC}}
=
\left\langle
\frac{|a_{\mathrm{out}}(t)|^2}{|a_{\mathrm{in}}|^2}
\right\rangle
=
\frac{\left|a_{\mathrm{in}}-\sqrt{\kappa_e}\,a_0\right|^2
+\kappa_e\sum_{n\neq0}|a_n|^2}{|a_{\mathrm{in}}|^2}.
\label{eq:Tdc}
\end{equation}
The quasi-DC transmission spectrum $T_{\mathrm{DC}}(\lambda)$ is then obtained by evaluating
Eq.~\eqref{eq:Tdc} as a function of laser detuning $\Delta(\lambda)$, as shown in Fig.~\ref{fig:AOM_power}b.

\section{Materials overview}
The performance of piezo-optomechanical transducers is ultimately dictated by the material
 on which they are realized. Material properties determine the strength of both the
electromechanical and optomechanical interactions that underpin coherent
microwave-optical transduction. In this section, we compare several widely used
piezoelectric material with the aim of assessing their suitability for efficient piezo-optomechanical devices.
\begin{figure}[!h]
    \centering
    \includegraphics[width=0.7
    \textwidth]{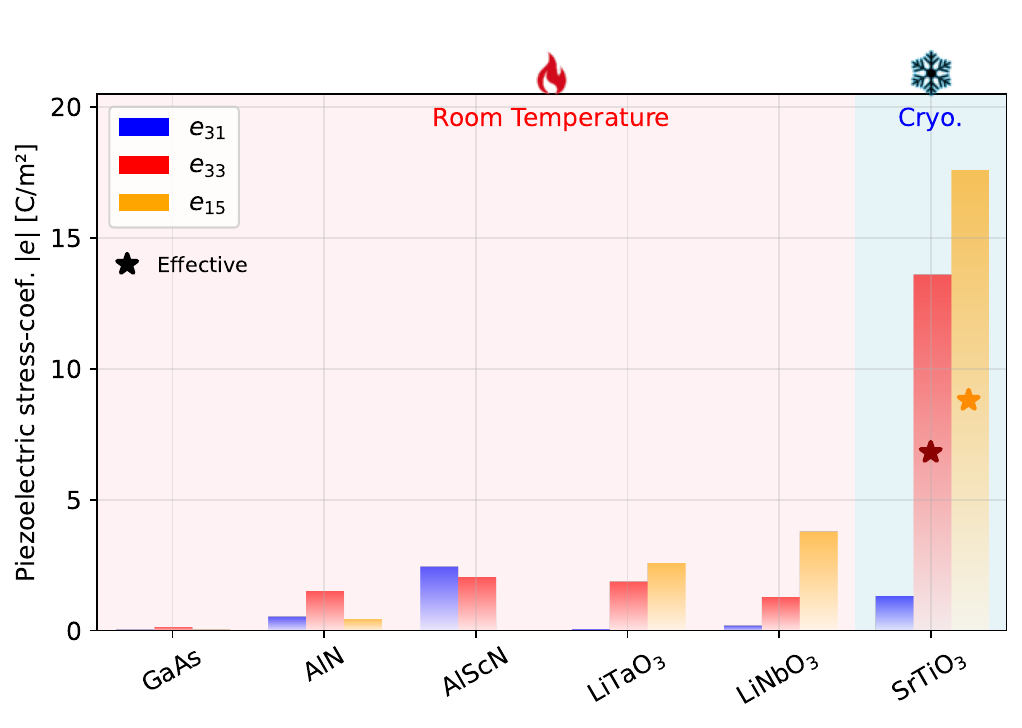}
    \caption{\textbf{Piezoelectric tensor comparison.} Piezoelectric tensor elements of commonly used materials. Effective parameters in SrTiO$_3$,
     arising from twin domain boundaries, are denoted by stars. \cite{1534782,electronics11142167,10.1063/1.4788728,SHUR2010204,Kushibiki1999-pt,Shur2011}.}
    \label{fig:piezo_coeffs}
\end{figure}
From an electromechanical perspective, a central figure of merit is the electromechanical
coupling rate $g_{\mathrm{EM}}$. In piezoelectric resonators, $g_{\mathrm{EM}}$ scales with the
square root of the effective electromechanical coupling coefficient $k^2$\cite{PhysRevApplied.13.014027}, which is
determined by the relevant piezoelectric tensor elements $e_{ij}$, as well as by device
geometry and modal overlap. Materials with large intrinsic piezoelectric coefficients, such
as AlN and LiNbO$_3$, naturally support large $k^2$ and can therefore achieve strong
electromechanical coupling in compact geometries. In contrast, GaAs exhibits a
comparatively weak piezoelectric response, resulting in smaller $k^2$ and correspondingly
reduced $g_{\mathrm{EM}}$, unless additional resonant enhancement or impedance-matching
strategies are employed. Fig.~\ref{fig:piezo_coeffs} compares the dominant piezoelectric
tensor elements for the materials considered here.

At cryogenic temperatures, the piezoelectric response of ferroelectric materials generally
decreases as the operating temperature moves further away from the Curie point. The
magnitude of this reduction is material dependent; for example, LiNbO$_3$ exhibits a
relatively modest decrease of approximately 6\% at 1.3~K \cite{Bukhari_2014}.

Optomechanical performance, by contrast, is primarily characterized by the single-photon
optomechanical coupling rate $g_0$, which depends on the photoelastic tensor, the refractive
index, and the achievable optical mode confinement. While conventional piezoelectric
materials provide moderate photoelastic coupling, hybrid photonic platforms enable access
to materials with substantially larger photoelastic coefficients. In particular,
SiN-on-SrTiO$_3$ and all-SrTiO$_3$ waveguides would leverage a strong photoelastic response
enabling enhanced optomechanical interaction strengths.
Table~\ref{tab:photoelastic_coeffs} summarizes the relevant photoelastic tensor elements for
representative optical guiding materials.

\begin{table}[!h]
\centering
\caption{Photoelastic tensor elements of commonly used optical guiding materials, where bold numbers indicate the high values  \cite{HUANG20031615,Tian:24,Zhu:21,BTO}.}
\label{tab:photoelastic_coeffs}
\begin{tabular}{ll}
\toprule
Material & Selected photoelastic tensor elements \\
\midrule
Si &
\( \mathbf{p_{11} = -0.094},\; p_{12} = 0.017,\; p_{44} = -0.051 \) \\

SiN &
\( \mathbf{ \left| p_{44}\right| \approx 0.086},p_{11} = 0.047\) \\

LiNbO\(_3\) &                                                 
\( \mathbf{p_{31} = 0.179},\; p_{33} = 0.071,\; p_{44} = 0.146,\; p_{41}=-0.151 , \dots\) \\

BaTiO\(_3\) &
\( \mathbf{p_{44} = 1.00}, \; p_{11} = 0.50,\; p_{33} = 0.77,\;
   p_{12} = 0.106,\; p_{13} = 0.20, \dots\) \\

SrTiO\(_3\) &
\( p_{\mathrm{eff}} \approx 0.5625 \) \\

\bottomrule
\end{tabular}
\end{table}

\bibliographystyle{sciencemag}
\bibliography{references}

\begin{thebibliography}{99}

\bibitem[24]{elastic1}
R.~O. Bell and G.~Rupprecht,
Elastic constants of strontium titanate,
\textit{Phys. Rev.} \textbf{129}, 90--94 (1963).
\href{https://doi.org/10.1103/PhysRev.129.90}{doi:10.1103/PhysRev.129.90}

\bibitem[25]{elastic2}
Scott, J.~F. and Ledbetter, H.,
Interpretation of elastic anomalies in SrTiO$_3$ at 37 K,
\textit{Zeitschrift für Physik B: Condensed Matter} \textbf{104}, 635--639 (1997).

\bibitem[15]{Tian:24}
H. Tian, J. Liu, A. Attanasio, A. Siddharth, T. Bl\'{e}sin, R. N. Wang,
A. Voloshin, G. Lihachev, J. Riemensberger, S. E. Kenning, Y. Tian,
T. H. Chang, A. Bancora, V. Snigirev, V. Shadymov, T. J. Kippenberg,
and S. A. Bhave,
Piezoelectric actuation for integrated photonics,
\textit{Adv. Opt. Photon.} \textbf{16}, 749--867 (2024).
\href{https://doi.org/10.1364/AOP.529288}{doi:10.1364/AOP.529288}

\end{thebibliography}

\end{document}